\documentclass{article}

\usepackage{arxiv}

\usepackage[utf8]{inputenc} 
\usepackage[T1]{fontenc}    
\usepackage{hyperref}       
\usepackage{url}            
\usepackage{booktabs}       
\usepackage{amsfonts}       
\usepackage{nicefrac}       
\usepackage{microtype}      
\usepackage{lipsum}		
\usepackage{graphicx}
\usepackage{natbib}
\usepackage{amsmath}
\usepackage{amsthm}
\usepackage{xcolor}
\usepackage{doi}
\usepackage[doublespacing]{setspace}
\usepackage[fontsize=12pt]{fontsize}
\usepackage{algorithm2e}
\usepackage{listings}
\usepackage{hyperref}
\usepackage{bbm}
\usepackage{crop}
\usepackage{graphicx}
\usepackage{caption}
\usepackage{amsmath}
\usepackage{array}
\usepackage{color}
\usepackage{xcolor}
\usepackage{amssymb}
\usepackage{flushend}
\usepackage{stfloats}
\usepackage[figuresright]{rotating}
\usepackage{chngpage}
\usepackage{totcount}
\usepackage{fix-cm}
\usepackage[T1]{fontenc}
\usepackage{lstbayes}
\DeclareMathOperator{\diag}{diag}

\newtheorem{result}{Result}
\SetKw{KwBy}{by}

\definecolor{comcolor}{rgb}{0,0.5,0}
\definecolor{funcolor}{rgb}{0,0.4,1}
\definecolor{concolor}{rgb}{0,0,1}
\definecolor{codegreen}{rgb}{0,0.6,0}
\definecolor{codegray}{rgb}{0.5,0.5,0.5}
\definecolor{codepurple}{rgb}{0.58,0,0.82}
\definecolor{backcolour}{rgb}{0.93,0.93,0.93}

\lstdefinestyle{mystyle}{
    backgroundcolor=\color{backcolour},   
    commentstyle=\color{codegreen},
    keywordstyle=\color{funcolor},
    numberstyle=\tiny\color{codegray},
    stringstyle=\color{codepurple},
    basicstyle=\ttfamily\footnotesize,
    breakatwhitespace=false,         
    breaklines=false,                 
    captionpos=b,                    
    keepspaces=true,                    
    numbersep=5pt,                  
    showspaces=false,                
    showstringspaces=false,
    showtabs=false,                  
    tabsize=2
}

\title{Bayesian Multivariate Sparse Functional Principal Components Analysis}

\author{ \href{https://orcid.org/0000-0001-9081-1955}{\includegraphics[scale=0.06]{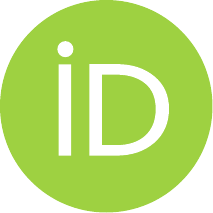}\hspace{1mm}Joseph Sartini}\thanks{Author webpage at \url{https://jsartini.github.io/Sartini-Stats/}} \\
	Department of Biostatistics\\
	Johns Hopkins University\\
	Baltimore, Maryland \\
	\texttt{jsartin1@jh.edu} \\
	\And
	Scott Zeger \\
	Department of Biostatistics\\
	Johns Hopkins University\\
	Baltimore, Maryland \\
    \And 
	Ciprian Crainiceanu \\
	Department of Biostatistics\\
	Johns Hopkins University\\
	Baltimore, Maryland \\
}

\renewcommand{\shorttitle}{Bayesian Multivariate Sparse FPCA}

\begin{document}
\maketitle

\singlespacing

\begin{abstract}
Functional Principal Components Analysis (FPCA) provides a parsimonious, semi-parametric model for multivariate, sparsely-observed functional data. Frequentist FPCA approaches estimate principal components (PCs) from the data, then condition on these estimates in subsequent analyses. As an alternative, we propose a fully-Bayesian inferential framework for multivariate, sparse functional data (MSFAST) which explicitly models the PCs and incorporates their uncertainty. MSFAST builds upon the FAST approach to FPCA for univariate, densely-observed functional data. Like FAST, MSFAST represents PCs using orthonormal splines and samples the orthonormal spline coefficients using parameter expansion. MSFAST extends FAST to multivariate, sparsely-observed data by (1) standardizing each functional covariate to mitigate poor posterior conditioning due to disparate scales; (2) using a better-suited orthogonal spline basis; (3) updating parameterizations for computational stability; (4) introducing routines that leverage multiple cores and threads to accelerate compute; (5) using a Procrustes-based posterior PC alignment procedure; and (6) providing efficient prediction routines. We evaluate MSFAST alongside existing implementations using simulations. MSFAST produces uniquely valid inferences and accurate estimates, particularly in smaller signal-to-noise regimes. MSFAST is motivated by and applied to a study of child growth, with an accompanying vignette illustrating the implementation step-by-step.
\end{abstract}

\onehalfspacing

\section{Introduction}
Sparse functional data has two key features: (1) observations are irregularly spaced; and (2) there are relatively few observations per subject \citep{crainiceanu2024book}. This type of data is collected in a wide range of applications, from biomarkers collected at clinic visits to physical functioning tests within prospective cohorts. While computational needs may differ from dense data, Functional Principal Components Analysis (FPCA) remains a powerful statistical tool which can perform dimensionality reduction, imputation, and prediction for sparse functional data \citep{yao_functional_2005, xiao_fast_2018}.

There are several approaches to sparse data FPCA, including: kernel smoothing \citep{Staniswalis1998}, local polynomial smoothing \citep{yao_functional_2005}, and penalized sandwich covariance smoothing \citep{xiao_fast_2018}. For more details see Chapter 3 of \cite{crainiceanu2024book}. Methods for multivariate sparse FPCA have also been proposed \citep{happ_multivariate_2018,li_fast_2020}. These methods estimate the functional principal components (FPCs) and then condition on them, treating them as fixed in subsequent inferences. This ignores the FPC estimation uncertainty, which can be large in smaller data sets or for components with smaller signals. As discussed by \citep{goldsmith_PCA_2013}, ignoring this uncertainty can lead to underestimation of FPC variability and over-confidence in estimates of both population-level functions and subject-level predictions. The variability of the FPC scores is also underestimated, which could lead to bias in downstream analyses. 

To account for estimation uncertainty in the FPCs, several fully Bayesian approaches to univariate sparse data FPCA have been proposed \citep{gertheiss_note_2017, ye_functional_2024}. Only \cite{nolan_efficient_2025} address multivariate, sparse data FPCA using a variational message passing approach, though \cite{jiang_bayesian_2022} implements a fully- Bayesian approach to multivariate, sparse longitudinal data analysis which models random effects semi-parametrically using correlated univariate FPCA models. 

In this paper, we introduce MSFAST: a fully Bayesian FPCA framework for multivariate, sparse functional data. MSFAST extends the FAST approach for univariate, densely-observed functional data \citep{sartini_fast_2026}. Similar to FAST, MSFAST combines orthonormal spline basis expansion of the principal components and efficient sampling of FPC spline coefficients using parameter expansion. MSFAST requires a series of innovations to appropriately handle multivariate, sparse data: (1) standardizing each functional covariate to prevent extreme covariate scales from causing numerical instability or improper smoothing; (2) changing the orthogonal spline basis used to model the FPCs to orthogonalized B-splines, which can better handle gaps in the observed data; (3) moving to non-central parameterizations for computational stability; (4) leveraging parallel compute by introducing additional sampling chains or multi-threaded likelihood computations; (5) introducing a Procrustes-based posterior FPC alignment procedure; and (6) developing efficient dynamic and static prediction methods.

This paper is organized as follows. Section~\ref{sec:MSFAST_methods} describes the MSFAST Bayesian model, including pre- and postprocessing. Section~\ref{sec:STAN} outlines the STAN implementation. Section~\ref{sec:MSFAST_sim} compares MSFAST with existing approaches in a simulation study. Section~\ref{sec:applications} includes a case study of child growth data. We conclude with a short discussion in Section~\ref{sec:MSFAST_disc}.

\section{Methods}\label{sec:MSFAST_methods}
 
\subsection{Sparse-data FPCA}\label{sec:Univariate}
We begin by building machinery in the univariate case, as much of the modeling infrastructure is directly transferable. Assume that functional data $Y_i(t_{ij})$ are observed at discrete time points $t_{ij}\in [0,1]$ for subjects $i = 1, \ldots, I$ and instances $j=1,\ldots,J_i$. The sampling points $t_{ij}$ can be different across subjects. Next, let $\mathbf{T}$ be the $M-$dimensional vector obtained by ordering the set $\{t_{ij}:i=1,\ldots,I,j=1,\ldots,J_i\}$. Note that $M$ does not always equal to $\sum_{i}J_i$ because some observations may be at the same sampling points.

Assume that $Y_i(t)$ follow the Gaussian functional principal components analysis (FPCA) model \citep{ramsaysilv2005, crainiceanu2024book} 
\begin{equation}
    Y_i(t) =  \mu(t) + \sum_{k = 1}^K \xi_{ik}\phi_k(t) + \epsilon_i(t)\;,\label{EQ:FPCA}
\end{equation}
where $\mu(t)$ is the population mean function; $\phi_k(t)$, $k=1,\ldots,K$, are the orthonormal FPCs; $\xi_{ik}\sim N(0, \lambda_k)$ where $\lambda_k$ are eigenvalues corresponding to $\phi_k(t)$; $\epsilon_i(t) \sim N(0, \sigma^2_\epsilon)$ are independent measurement errors; and $\xi_{ik}, \epsilon_i(t)$ are mutually independent over $i,k$. The number of FPCs $K$  is fixed to achieve a certain percent variance explained. This can be assessed by either over-parameterizing the model with large $K$ and testing various levels of truncation, or by performing sensitivity analysis over a range of $K$ values \citep{sartini_fast_2026}.
The likelihood for Model~\eqref{EQ:FPCA} conditional on the parameters is
\begin{equation}
    \prod_{i = 1}^I\prod_{j=1}^{J_i} N\{Y_i(t_{ij})|\mu(t_{ij}) + \sum_{k = 1}^K \xi_{ik}\phi_k(t_{ij}), \sigma^2_\epsilon\}\;,\label{EQ:Like_obs}
\end{equation}
where $N(y|\mu,\sigma^2)$ denotes the normal probability density function with mean $\mu$ and variance $\sigma^2$ evaluated at $y$. Frequentist approaches to sparse FPCA maximize Equation~\eqref{EQ:Like_obs} conditional on $\phi_k(t)$ estimates from covariance smoothing. To account for estimation uncertainty in the FPCs, we model them jointly with the other parameters \citep{sharpe_uncertainty_2016}. This is particularly relevant for sparse data, which contains less information about the underlying covariance structure \citep{goldsmith_PCA_2013}.

Only small adjustments are required to adapt FAST to sparsely observed functional data. We again represent the mean $\mu(t)$ and principal components $\phi_k(t)$ using a rich orthonormal spline basis, reducing the problem of modeling the infinite dimensional latent functions to considering a small dimensional space of spline coefficients. In particular, we consider $Q-$dimensional basis $\mathbf{B}(t) = \{B_1(t), \ldots, B_Q(t)\}$ such that the $B_q(t)$ functions are orthonormal with respect to the $L_2$ scalar product: $\langle B_q(t), B_{q'}(t)\rangle = \int_0^1 B_q(t) B_{q'}(t)dt$. This representation allows us to efficiently evaluate the functions $\phi_k(t), \mu(t)$ at any time-point $t$ by simply evaluating the spline bases. We choose $Q$ sufficiently large ($20$-$40$) as suggested by the penalized spline literature \citep{ruppert_selecting_2002}. We adjust the basis from the more localized Splinets \citep{liu_splinets_2020} to orthogonalized B-splines \citep{redd_comment_2012}.
 
As shown in \cite{sartini_fast_2026}, the FPCs $\phi_k(t) = \mathbf{B}(t)\psi_k$ are orthonormal if and only if the column-bound matrix $\boldsymbol{\Psi} = [\psi_1|\cdots|\psi_K]$ is orthonormal. That is, $\boldsymbol{\Psi} \in \mathcal{V}_{K,Q}$, the Stiefel manifold of orthonormal matrices with dimension $Q\times K$. To sample $\mathbf{\Psi} \in \mathcal{V}_{K,Q}$, we use the parameter expansion technique based on the polar decomposition proposed by \cite{jauch_monte_2021}. This technique involves sampling a latent matrix $\mathbf{X} \in \mathbb{R}^{Q \times K}$, performing polar decomposition, and taking the orthonormal component as the corresponding sample of $\boldsymbol{\Psi}$. When each entry in $\mathbf{X}$ has an independent standard normal prior distribution, this induces a uniform prior for $\boldsymbol{\Psi}$ over the manifold $\mathcal{V}_{K, Q}$ \citep{chikuse_statistics_2003}.
To calculate $\boldsymbol{\Psi}$, we obtain the eigendecomposition $\mathbf{X}^\top\mathbf{X} = \mathbf{ZDZ}^\top$ and define $\mathbf{\Psi}=\mathbf{XZD}^{-1/2}\mathbf{Z}^\top$. The transformation $\mathbf{X} \rightarrow \mathbf{\Psi}$ is unique as long as $\mathbf{X}$ has full column rank \citep{polar_source}, which ensures that the target density and corresponding gradients are proper \citep{betancourt_conceptual_2018}. 

As in FAST, we control the complexity of functional components $\mu(t)$ and $\phi_k(t)$ using smoothing penalty priors of the form $\alpha \int f^2(t)dt + (1-\alpha)\int \{f''(t)\}^2dt$ for generic $f(\cdot)$ and tuning parameter $\alpha = 0.1$. This penalty can be expressed as a quadratic form in the spline coefficients: $\psi_k^\top \mathbf{P}_\alpha \psi_k$ for $\phi_k(t)$ and $Q \times Q$ penalty matrix $\mathbf{P}_\alpha$ (see Supplement Section~\ref{supp:penalties}). Each penalty is scaled by a corresponding stochastic smoothing parameter $h_k$, which we assume to have Gamma prior with shape and rate hyper-parameters $\alpha_\psi$ and  $\beta_\psi$, respectively. Conditional on the $h_k$, we introduce a smoothing spline prior on the coefficients $\psi_k$ with the additional constraint of orthonormality. As demonstrated in prior work, the joint prior on the spline coefficients $\psi_k$ and smoothing parameters $h_k$ is proper for any valid choice of hyper-parameters $\alpha_\psi > 0$ and $\beta_\psi > 0$ \citep{sartini_2026_suffcond}.

\subsection{Multivariate, Sparse-data FPCA}\label{sec:Multivariate}

Model~\eqref{EQ:FPCA} can be extended to multivariate sparse functional data \citep{happ_multivariate_2018, li_fast_2020}. Denote the observed data as $Y_{i}^{(p)}(t_{ij}^{(p)})$, where the new index $p=1,\ldots,P$ indicates functional covariate. For notation simplicity, we consider all functional observations to be in time, but the proposed methods work for any combination of functional domains. Different functional variables are not necessarily observed at the same time points within and between subjects. Denote by $\mathbf{T}$ the ordered union of all observed points, and let $\mathbf{T}_i^{(p)} \subseteq \mathbf{T}$ be the set of points where $Y_i^{(p)}(t)$ is observed (with $J_i^{(p)} = |\mathbf{T}_i^{(p)}|$). We use $\mathbf{Y}_i(t) = \{Y_i^{(1)}(t), \ldots, Y_i^{(P)}(t)\}$ to denote the $P$-dimensional vector of functions for subject $i$ at time $t$. \cite{happ_multivariate_2018} showed that $\mathbf{Y}_i(t)$ can be decomposed according to the Kosambi-Karhunen-Loève Theorem \citep{kosambi_statistics_1943, karhunen_uber_1947, loeve_probability_1978}:
\begin{equation}
    \mathbf{Y}_i(t) = \boldsymbol{\mu}(t) + \sum_{k = 1}^{\infty}\xi_{ik}\boldsymbol{\phi}_k(t) \approx \boldsymbol{\mu}(t) + \sum_{k = 1}^K \xi_{ik}\boldsymbol{\phi}_k(t) + \boldsymbol{\epsilon}_i(t)\;,
    \label{EQ:Multivariate_FPCA}
\end{equation}
where $\boldsymbol{\mu}(t) = \{\mu^{(1)}(t), \ldots, \mu^{(P)}(t)\}^\top$,  $\boldsymbol{\phi}_k(t) = \{\phi_k^{(1)}(t), \ldots, \phi_k^{(P)}(t)\}^\top$, and  $\boldsymbol{\epsilon}_i(t) \sim {\rm MVN}(\mathbf{0}_P, \boldsymbol{\Sigma}_\epsilon)$ for $\boldsymbol{\Sigma}_\epsilon = \text{diag}(\sigma_1^2, \ldots, \sigma_P^2)$. Each functional variable has a unique mean and measurement error variance, the eigenfunctions $\boldsymbol{\phi}_k(t)$ are obtained by concatenating the corresponding components specific to each functional covariate, and the scores $\xi_{ik}$ are unique for each subject $i$ and eigenfunction $k$ -- not differing over functional variable $p$. The multivariate eigenfunctions $\boldsymbol{\phi}_k(t)$ are orthonormal with respect to the sum inner product: $\langle \boldsymbol{\phi}_j(t), \boldsymbol{\phi}_k(t) \rangle = \sum_{p = 1}^P \int_0^1 \phi_j^{(p)}(t) \phi_k^{(p)}(t) dt$.

The corresponding multivariate likelihood is a direct extension of the univariate likelihood in Equation~\eqref{EQ:Like_obs}, with conditionally-independent residuals both over time and between covariates
\begin{equation}
    \prod_{p = 1}^P \prod_{i = 1}^I {\rm MVN}(Y_i^{(p)}(\mathbf{T}_i^{(p)})|\mu^{(p)}(\mathbf{T}_i^{(p)}) + \sum_{k = 1}^K\xi_{ik} \phi_k^{(p)}(\mathbf{T}_i^{(p)}), \sigma^2_p \mathbf{I}_{J_i^{(p)}}) \;.\label{EQ:MV_Likelihood}
\end{equation}
As in the univariate case (Section~\ref{sec:Univariate}), we represent the functional components $\mu^{(p)}(t)$, $\phi_k^{(p)}(t)$ for all $p$ and $k$ using a $Q-$dimensional spline basis $\mathbf{B}(t)$ which is orthonormal with respect to the $L_2$ scalar product. In particular, we let $\mu^{(p)}(t) = \mathbf{B}(t)w_\mu^{(p)}$ and $\phi_k^{(p)}(t) = \mathbf{B}(t) \psi_k^{(p)}$ for all $p,k$, where $w_\mu^{(p)}, \psi_k^{(p)} \in \mathbb{R}^Q$. Basis dimension $Q$ is still chosen between 20-40 based upon penalized spline literature, and sensitivity analyses indicate that MSFAST is robust to this choice so long as $Q$ is not too small (see Supplement~\ref{supp:q_sens}).

With this notation $\boldsymbol{\mu}(t) = [\mathbf{I}_P \otimes \mathbf{B}(t)]\mathbf{w}_\mu$ and $\boldsymbol{\phi}_k(t) = [\mathbf{I}_P \otimes \mathbf{B}(t)]\boldsymbol{\psi}_k$, for the $PQ-$dimensional concatenated spline parameters $\mathbf{w}_\mu = \{w_\mu^{(1)}, \ldots, w_\mu^{(P)}\}^\top$ and $\boldsymbol{\psi}_k = \{\psi_k^{(1)}, \ldots, \psi_k^{(P)}\}^\top$, respectively. Here $\mathbf{I}_P$ is the $P \times P$ identity matrix and $\otimes$ is the Kronecker product. We can evaluate $\mathbf{B}(t)$ at any desired $t \in \mathbf{T}$ and apply the above to evaluate all functional components at that point.

We next demonstrate that the FPCs are orthonormal with respect to the sum inner product if and only if the column-bound $PQ\times K$ dimensional matrix of FPC spline coefficients $\boldsymbol{\Psi} = [\boldsymbol{\psi}_1|\cdots|\boldsymbol{\psi}_K]$ is orthonormal, i.e. $\boldsymbol{\Psi}^\top \boldsymbol{\Psi} = \mathbf{I}_K$.
\begin{align*}
    \langle \boldsymbol{\phi}_k(t), \boldsymbol{\phi}_{k'}(t) \rangle & = \sum_{p = 1}^P \int_0^1 \phi_k^{(p)}(t) \phi_{k'}^{(p)}(t)dt \\
    & = \sum_{p = 1}^P \int_0^1 \mathbf{B}(t)\psi_k^{(p)} \mathbf{B}(t)\psi_{k'}^{(p)}dt \\
    & = \sum_{p = 1}^P \sum_{i = 1}^Q \sum_{j = 1}^Q \psi_{k,i}^{(p)}\psi_{k',j}^{(p)} [\int_0^1 \mathbf{B}_i(t) \mathbf{B}_j(t)dt]\\
    & = \sum_{p = 1}^P (\psi_{k}^{(p)})^\top \psi_{k'}^{(p)} = \boldsymbol{\psi}_k^\top \boldsymbol{\psi}_{k'}\;.
\end{align*}
Therefore, $\langle \boldsymbol{\phi}_k(t), \boldsymbol{\phi}_{k'}(t) \rangle = \boldsymbol{\psi}_k^\top \boldsymbol{\psi}_{k'}$, from which it follows that the functions $\boldsymbol{\phi}_k(t)$ are orthonormal if and only if the matrix $\boldsymbol{\Psi}$ is orthonormal. 

As sampling the FPCs reduces to sampling the orthonormal matrix $\mathbf{\Psi}$ of FPC spline weights, we can again use the parameter expansion based upon polar decomposition proposed by \cite{jauch_monte_2021} to sample this matrix. The only aspect of this approach which is not simple matrix multiplication is eigendecomposing $\mathbf{X}^\top \mathbf{X}$, which scales with the number of principal components $K$. This is why using polar decomposition to sample $\boldsymbol{\Psi}$ remains efficient as $P,Q$ scale. As in the univariate case, we use standard normal priors on the entries of $\mathbf{X}$ to induce a uniform prior distribution over the manifold $\mathcal{V}_{K, PQ}$ for the coefficient matrix $\boldsymbol{\Psi}$.

We control complexity of the functional components $\mu^{(p)}(t), \phi^{(p)}_k(t)$ using smoothing spline priors as in the univariate case. The penalty matrix $\mathbf{P}_\alpha$ remains the same as in Subsection~\ref{sec:Univariate}, with $\alpha = 0.1$ (though MSFAST is robust to this choice, see Supplement Section~\ref{supp:alpha_sens}). In extension to multivariate outcomes, smoothing parameters $h_k$ are unique to each eigenfunction but are shared across variates. We again employ Gamma priors with shape and rate hyper-parameters $\alpha_\psi, \beta_\psi$, respectively. Result~\ref{rmk:proper_prior} indicates that, as in the univariate case, the joint prior on the eigenfunctions and their smoothing parameters is always proper and results in a well-defined posterior. Proof can be found in Supplement Section~\ref{supp:proper_prior}.

\begin{result}\label{rmk:proper_prior}
    The joint prior on the eigenfunction spline weights $\{\boldsymbol{\psi}_1, \ldots \boldsymbol{\psi}_K\}$ and smoothing parameters $\{h_1, \ldots, h_K\}$ is proper for any valid choices of $\alpha_\psi  > 0$ and $\beta_\psi > 0$.
\end{result}

Our inference is based on fixing the number of FPCs $K$ and quantifying the proportion of joint variance explained as a function of $K$. In practice, we can choose a large initial $K$ and truncate to the smallest basis dimension that explains a given proportion of data variance (e.g., $95$\%). Sensitivity analyses indicate that, even when the number of FPCs $K$ is mis-specified, MSFAST still well-recovers the model components (see Supplement Section~\ref{supp:k_sens}). It is important to note that MSFAST automatically produces global and covariate-specific variance explained metrics. \cite{golovkine_estimation_2025} point out that choosing $K$ based on the variance explained for each covariate does not generally correspond to joint variance explained.

While notation is more complex in the multivariate case, ideas from univariate sparse functional model extend directly to multivariate sparse data, and the {\ttfamily STAN} implementation is straightforward as discussed in Section~\ref{sec:STAN}.

\subsection{MSFAST Updates to Handle Multivariate, Sparse Data}\label{sec:changes}

FAST is a fully Bayesian method for dense functional data that jointly models the FPCs with all other model parameters to account for their uncertainty. Here, we focus on methods for sparse functional data, which contains substantially less likelihood information, and on multivariate functional data, which introduces cross-covariances between functional variables to the model. These differences impact inference and require careful adjustments to pre-processing, the FAST Bayesian implementation, and post-processing to ensure valid inferences and computational feasibility.

During pre-processing, MSFAST standardizes each covariate by subtracting their respective sample means and dividing by their standard deviations. After fitting, parameters are mapped back to the original scale without loss of information. To understand the motivation for this step, consider the structure of Model~\eqref{EQ:Multivariate_FPCA}. Due to the sharing of scores across covariates, any difference in the scales of the functional variables must be handled by the FPCs. Linear scaling of the FPCs also scales their smoothness penalty (see Supplement Section~\ref{supp:penalties}). As the smoothing parameters are shared between variates, disparate scales would result in larger variates largely determining the smoothing parameters. Then, maintaining the proper degree of smoothing requires that variates be on the same scale. This also protects against numerical stability issues introduced by large disparities in covariate scales, which could cause eigenfunctions to approach 0 or 1. 

As noted in Section~\ref{sec:Univariate}, MSFAST uses orthonormalized B-splines \citep{redd_comment_2012} instead of the Splinets \citep{liu_splinets_2020} used by FAST. Splinet bases are designed to maintain B-spline locality after orthonormalization, such that the resulting functions are close to zero far from their knot locations. This is numerically useful for dense data, but is unstable for sparse data over intervals where data is not observed. So, MSFAST uses simple orthonormalized B-splines that are smoother and less local (see Supplement Section~\ref{supp:bases}).

MSFAST tunes the parameterizations and extends the data structures used by FAST to handle multivariate, sparse data. First, MSFAST uses non-centered parameterization for the normally distributed scores \citep{papaspiliopoulos_general_2007}. While this technique can introduce irregularities to the score distributions at the data scale, we mitigate these issues with the alignment routine described in Section~\ref{sec:ACE}. Second, MSFAST represents the sparse data as a single concatenated vector -- ordered by functional variate -- with corresponding labels for subject and domain location. Ordering by functional variate facilitates multi-threaded computation of likelihood contributions (see Supplement Section~\ref{supp:parallel}).

Finally, we introduce a new posterior sample alignment routine using the Procrustes transformation to reduce multi-modality due to low signal and/or similar eigenvalues; see Section~\ref{sec:ACE} for more details. We also introduce a new method designed to conduct dynamic prediction without the need to refit the model when new data is observed. The method substantially reduces computational times and works as well as complete model refitting for our application; see Section~\ref{sec:prediction} for details.

\subsection{Estimation, Alignment, and Convergence} \label{sec:ACE}
To efficiently form a final posterior FPC estimate which is $L^2$ orthonormal, we work on the spline coefficient space and transform to the functional space. First, consider the random deviation from the mean for arbitrary subject $i$ according to Model~\ref{EQ:Multivariate_FPCA}: $\mathbf{U}_i(t) = \sum_{k = 1}^{K}\xi_{ik}\boldsymbol{\phi}_k(t)$. We can represent each $\mathbf{U}_i(t)$ using spline expansion: $\mathbf{U}_i(t) = [\mathbf{I}_P \otimes \mathbf{B}(t)]\sum_{k = 1}^{K} \xi_{ik}\boldsymbol{\psi}_k$, where $\sum_{k = 1}^{K} \xi_{ik}\boldsymbol{\psi}_k = L_i \in \mathbb{R}^{Q}$ is the representation of $\mathbf{U}_i(t)$ on the spline coefficient space. As we consider only those functions within the span of our spline basis, and that basis is orthonormal, the FPCs of the $U_i(t)$ correspond to the right singular vectors of the matrix of latent representations $\mathbf{L} = [L_1, \ldots, L_I]^\top \in \mathbb{R}^{N \times Q}$. We calculate the posterior mean of $\mathbf{L}$, perform singular value decomposition (SVD), and take the right singular vectors as $\widehat{\boldsymbol{\Psi}}$. We then calculate our posterior FPC estimate as $\widehat{\boldsymbol{\phi}}_k(t) = [\mathbf{I}_P \otimes \mathbf{B}(\mathbf{T})]\widehat{\boldsymbol{\Psi}}$. All operations are on the $PQ-$dimensional spline space, maintaining computational efficiency.

The posterior alignment procedure for MSFAST starts by obtaining the FPC matrix from each sample $s$ evaluated at all times $t \in \mathbf{T}$, denoted $\boldsymbol{\Phi}^{(s)} = [\boldsymbol{\phi}^{(s)}_1(\mathbf{T})|\cdots |\boldsymbol{\phi}^{(s)}_K(\mathbf{T})] = [\mathbf{I}_P \otimes \mathbf{B}(\mathbf{T})]\boldsymbol{\Psi}^{(s)} \in \mathbb{R}^{PM \times K}$. We then identify the optimal rotation matrices, $\mathbf{R}^{(s)} \in \mathbb{R}^{K \times K}$, which register each $\boldsymbol{\Phi}^{(s)}$ to a fixed reference matrix, $\widetilde{\boldsymbol{\Phi}}$. The $\mathbf{R}^{(s)}$ are chosen to minimize the Frobenius norm $||\widetilde{\boldsymbol{\Phi}} - \boldsymbol{\Phi}^{(s)}\mathbf{R}^{(s)}||_F$. This is a classical application of Procrustes analysis \citep{Trendafilov2021} and has the closed form solution $\mathbf{R}^{(s)} = \mathbf{U}^{(s)}\{\mathbf{V}^{(s)}\}^\top$, where the matrices $\mathbf{U}^{(s)}$ and $\mathbf{V}^{(s)}$ are obtained from the Singular Value Decomposition (SVD): $\{\boldsymbol{\Phi}^{(s)}\}^\top \widetilde{\boldsymbol{\Phi}} = \mathbf{U}^{(s)}\mathbf{D}^{(s)}\{\mathbf{V}^{(s)}\}^\top$ \citep{watson_solution_1994, MatComp_1989}. As both $\mathbf{U}^{(s)}$ and $\mathbf{V}^{(s)}$ are unitary, it follows that $\mathbf{R}^{(s)}\{\mathbf{R}^{(s)}\}^\top = \mathbf{I}_K$. Therefore, the posterior samples of the scores $\tilde{\xi}^{(s)}_{i}$ corresponding to the rotated FPC bases can be obtained as $\tilde{\xi}^{(s)}_{i} = \{\mathbf{R}^{(s)}\}^\top\xi^{(s)}_i$, for original scores $\xi^{(s)}_i \in \mathbb{R}^K$.

Any estimate of the true $\boldsymbol{\Phi}$ can be used as the reference point $\widetilde{\boldsymbol{\Phi}}$. We can use the MSFAST estimates or leverage suitably-scaled estimates from \cite{xiao_fast_2018}. After alignment, convergence of all model parameters can be assessed using R-hat statistics \citep{sartini_fast_2026, gelman_inference_1992}. We find that the posterior alignment procedure improves convergence properties, substantially shortening necessary warm-up (as little as 250 iterations were sufficient in simulation).

\subsection{Imputation and Dynamic Prediction}\label{sec:prediction}

Prediction refers to obtaining estimators $\widehat{\mathbf{Y}}_i(t)$ and their associated uncertainty based on Model~\eqref{EQ:Multivariate_FPCA} at any time $t$. In practice, predictions are conducted between the minimum and maximum of the observed times, $t\in[\min(\mathbf{T}),\max(\mathbf{T})]$, including at points where the data are observed for subject $i$. Prediction can be conducted after all data are collected for all study subjects (static prediction/imputation) or when all data are collected for some of the subjects while data are being observed sequentially for other study subjects (dynamic prediction) \citep{ivanescu_dynamic_2017, ivanescu_outlier_2024}. 

Conducting static prediction is straightforward using MSFAST.  The model produces posterior samples of the spline coefficients $\boldsymbol{\Psi}$, mean parameters $\mathbf{w}_\mu$, and FPC scores. Samples from the joint posterior distribution of latent trajectories at any $t$ can be obtained using functions of the sampled model parameters: $\mathbf{Y}_i(t) = [\mathbf{I}_P \otimes \mathbf{B}(t)]\{\mathbf{w}_\mu + \sum_{k = 1}^K \xi_{ik} \boldsymbol{\psi}_k(t)\}$ for any time $t$. 

Dynamic prediction can be conducted by refitting the model repeatedly with updated data for the subject of interest. This is easy to implement, but is more computationally expensive. A faster alternative is to fit the model to the available data and make the assumption that the new subject's data does not substantially influence the $\boldsymbol{\Psi}$/$\mathbf{w}_\mu$ posterior distributions. Under this assumption, the posterior distribution of the new subject's scores, conditional upon their data and the population parameters, is a multivariate normal with mean and variance given in Result~\ref{Thm:Scores}. This conditional posterior, along with those for all other sampled model parameters, is derived in Supplement Section~\ref{supp:posteriors}.

\begin{result} \label{Thm:Scores}
The conditional posterior distributions of the scores $\xi_{i}$ for subject $i$ are multivariate normal $[\xi_{i}|{\rm others}] \sim MVN(\mathbf{M}_i, \boldsymbol{\Sigma}_i)$, where
\begin{align*}\mathbf{M}_i & = \boldsymbol{\Sigma}_i\left[\sum_{p = 1}^P\frac{ (\mathbf{R}_i^{(p)})^\top \mathbf{B}(\mathbf{T}_i^{(p)}) \boldsymbol{\Psi}^{(p)}}{\sigma_p^2}\right]^\top,\\
\boldsymbol{\Sigma}^{-1}_i & = \sum_{p = 1}^P\frac{(\boldsymbol{\Psi}^{(p)})^\top \mathbf{B} (\mathbf{T}_i^{(p)})^\top \mathbf{B}(\mathbf{T}_i^{(p)}) \boldsymbol{\Psi}^{(p)}}{\sigma_p^2} + \boldsymbol{\Lambda}^{-1},
\end{align*}
$\mathbf{B}(\mathbf{T}_i^{(p)})$ is the $J_i^{(p)} \times Q$ dimensional matrix of basis functions evaluated at the time points in $\mathbf{T}_i^{(p)}$, $\boldsymbol{\Psi}^{(p)} = [\psi_1^{(p)}|\cdots|\psi_K^{(p)}] \in \mathbb{R}^{Q \times K}$, $\boldsymbol{\Lambda}$ is the diagonal matrix of eigenvalues $\lambda_k$, and $\mathbf{R}_i^{(p)} = Y_i^{(p)}(\mathbf{T}_i^{(p)}) - \mathbf{B}(\mathbf{T}_i^{(p)}) \mathbf{w}_\mu^{(p)}$ is the $J_i^{(p)}$ dimensional residual.
\end{result}
Under the assumption that the posterior distributions of $\mathbf{w}_\mu$ and $\boldsymbol{\Psi}$ are not substantially affected by the data from subject $i$, the scores $\xi_i$ can be sampled directly from this multivariate normal distribution. This is much faster than refitting the model every time new data becomes available. For greater detail, see Supplement Section~\ref{supp:cond_pred}.

\section{Implementation in STAN}\label{sec:STAN} 

We now provide the {\ttfamily STAN} \citep{carpenter_stan_2017} implementation of MSFAST. First, the {\ttfamily data} section contains the observed data {\ttfamily Y} in a stacked vector accompanied by the indexing arrays {\ttfamily Subj} and {\ttfamily S}. These arrays indicate the corresponding subjects ($i$ in statistical notation) and sampling points (index in observation times $\mathbf{T}$). The data is stacked by covariate, with {\ttfamily Tp\_cardin} indicating the number of entries corresponding to each sequential variate ($|\mathbf{T}^{(p)}|$ in statistical notation). This sparse data structure allows calculating the likelihood  without computing latent trajectories $Y_i(t)$ at all locations $t \in \mathbf{T}$. The value {\ttfamily B} encodes the basis matrix evaluated at all time points: $\mathbf{B}(\mathbf{T}) \in \mathbb{R}^{M \times Q}$, and {\ttfamily P\_alpha} encodes the penalty matrix $\mathbf{P}_\alpha$ associated with basis $\mathbf{B}(t)$.

As mentioned in Sections~\ref{sec:Univariate} and ~\ref{sec:Multivariate}, we fix the number of FPCs {\ttfamily K} and the spline basis dimension {\ttfamily Q}. Selection of {\ttfamily K} is an important and difficult problem, but is not the scope of this work. Here we obtain the percent variability explained by each  {\ttfamily K} and fix a particular level, though inference can be provided for any level of explained variability. The choice of {\ttfamily Q} is more straightforward, as it need only be sufficiently large (e.g., $20$ to $40$) to capture the maximum complexity of the mean and principal components. The smoothness penalties control the complexity of estimators. 

\singlespacing
\begin{lstlisting}[language=Stan]
data {
  int N;   // Number of time series
  int M;   // Cardinality of observation time set
  int L;   // Total number of observed data points
  int Q;   // Spline basis dimension
  int K;   // FPC basis dimension
  int P;   // Number of covariates
  
  vector[L] Y;                // Concatenated observations
  array[L] int<upper=N> Subj; // Subject indices
  array[L] int<upper=M> S;    // Time/domain indices
  array[P] int Tp_card;       // Data points per covariate
  
  matrix[M, Q] B;           // Orthogonal spline basis
  matrix[Q, Q] P_alpha;     // Spline penalty matrix
}
\end{lstlisting}
\doublespacing

In the {\ttfamily parameters} block, we declare the eigenvalues {\ttfamily lambda} and unique smoothing parameters for each FPC in {\ttfamily H} and mean function in {\ttfamily h\_mu}. Finally, the {\ttfamily Scores\_Raw} matrix corresponds to the matrix of unscaled FPC scores, with $(i,k)$ entry equaling $\xi_{ik}/\sqrt{\lambda_k}$.

\singlespacing
\begin{lstlisting}[language=Stan]
parameters {
  vector<lower=0>[P] sigma2; // Error in observation
  
  // Fixed-effects
  vector[P*Q] w_mu;         // Mean spline coefficients
  vector<lower=0>[P] h_mu;  // Mean smoothing parameters
  
  // Covariance structure
  vector<lower=0>[K] lambda;      // Eigenvalues
  vector<lower=0>[K] H;           // EF Smoothing parameters
  matrix[P*Q, K] X;               // Unconstrained FPC weights
  matrix[N, K] Scores_Raw;        // EF scores with scale 1
}
\end{lstlisting}
\doublespacing

The key component of the {\ttfamily transformed parameters} block is the definition of $\mathbf{\Psi}$ ({\ttfamily Psi}), the orthonormal component of the polar decomposition of the unconstrained matrix $\mathbf{X}$. More specifically, $\mathbf{\Psi}=\mathbf{XZD}^{-1/2}\mathbf{Z}^\top$, where $\mathbf{Z}$ ({\ttfamily evec\_XtX}) is the $K\times K$ dimensional matrix containing the eigenvectors of $\mathbf{X}^\top\mathbf{X}$, and $\mathbf{D}$ ({\ttfamily diag(eval\_XtX)}) is a diagonal matrix containing the corresponding eigenvalues. Sampling the latent, unconstrained $\mathbf{X}$ ({\ttfamily X}) matrix is crucial, as Hamiltonian Monte Carlo is most effective in unconstrained spaces \citep{betancourt_conceptual_2018}. Further, the dimension of $\mathbf{X}^\top\mathbf{X}$ is constrained by the number of FPCs $K$, which is often modest as most functional data is well-represented with moderate $K$. We also scale {\ttfamily Scores\_raw} by the eigenvalues in this block to produce the {\ttfamily Scores} variable. This implements the non-central parameterization described in Section~\ref{sec:changes}. 

\singlespacing
\begin{lstlisting}[language=Stan]
transformed parameters{
  matrix[N, K] Scores;  // Scaled scores (Xi)
  matrix[P*Q, K] Psi;   // Orthonormal basis weights

  Scores = Scores_Raw * diag_matrix(sqrt(lambda));
  
  // Polar decomposition
  matrix[K,K] evec_XtX = eigenvectors_sym(crossprod(X)); 
  vector[K] eval_XtX = eigenvalues_sym(crossprod(X));
  Psi = X*evec_XtX*diag_matrix(1/sqrt(eval_XtX))*evec_XtX'; 
}
\end{lstlisting}
\doublespacing

The (inverse) variance component priors are specified first within the {\ttfamily model} block. Note that the smoothing parameters have prior shapes and rates of $0.01$, remaining at most weakly informative. All other variance components have standard, weakly informative priors. Then, we implement the smoothing penalties on the target posterior. Next comes the crucial independent $N(0,1)$ priors on {\ttfamily X}, which induce a uniform prior on $\mathbf{\Psi}$ ({\ttfamily Psi}) over the manifold $\mathcal{V}_{K, PQ}$ (as described in Section~\ref{sec:MSFAST_methods}). Sampling the smaller-dimensional, unconstrained matrix $\mathbf{X}$ is a major factor that ensures the scalability and robustness of MSFAST. We conclude with standard normal distributions for {\ttfamily Scores\_raw}, such that {\ttfamily Scores} will have the desired $\xi_{ik} \sim N(0, \lambda_k)$ distributions.

\singlespacing
\begin{lstlisting}[language=Stan]
model {
  // Variance component priors
  lambda (*$\sim$*) inv_gamma(0.001, 0.001); 
  sigma2 (*$\sim$*) inv_gamma(0.001, 0.001);
  
  // Smoothing priors
  h_mu (*$\sim$*) gamma(0.01, 0.01); 
  H (*$\sim$*) gamma(0.01, 0.01); 
  
  int sx, ex;
  for(p in 1:P){
    sx = (p-1)*Q+1;
    ex = p*Q;
    target += Q/2*log(h_mu[p])-h_mu[p]/2*quad_form(P_alpha, w_mu[sx:ex]);
    for(k in 1:K){
      target += Q/2*log(H[k])-H[k]/2*quad_form(P_alpha, Psi[sx:ex,k]);
    }
  }
  
  // Normal prior inducing uniform prior on the Stiefel manifold
  to_vector(X) (*$\sim$*) std_normal();
  
  // Priors on unscaled scores
  to_vector(Scores_Raw) (*$\sim$*) std_normal(); (*$\cdots$*)
\end{lstlisting}
\doublespacing

Next are the likelihood calculations, the key structural difference between MSFAST and FAST. We order the data provided to MSFAST by functional variable. This allows MSFAST to calculate the variable-specific likelihood contributions separately, making the task parallelizable (see Supplemental Section~\ref{supp:parallel}). Within each variable, we first sub-set all model components to the corresponding variate. Then, the variable-specific FPC matrix $\boldsymbol{\Phi}^{(p)} = [\phi^{(p)}_1(\mathbf{T})|\cdots|\phi^{(p)}_K(\mathbf{T})] \in \mathbb{R}^{M \times K}$ (denoted {\ttfamily Phi\_mat}) is calculated from the corresponding spline weights. {\ttfamily Phi\_mat} is then used to calculate the latent smooths evaluated at the observed time points, $\mu^{(p)}(\mathbf{T}_i^{(p)}) +\sum_{k = 1}^K \xi_{ik} \phi_k^{(p)}(\mathbf{T}_i^{(p)})$ properly concatenated. This quantity is referred to as {\ttfamily Theta}. We calculate {\ttfamily Theta} once per sample, use it to evaluate the variable-specific likelihood in Model~\eqref{EQ:Multivariate_FPCA}, then discard it to save memory. We re-use the integer variables {\ttfamily sx,ex} from the prior portion of the model block.

\singlespacing
\begin{lstlisting}[language=Stan]
model { 
  (*$\cdots$*)
  // Model likelihood
  int pos = 1;
  for(p in 1:P){
    // Declare
    matrix[M, K] Phi_mat;
    vector[M] Mu;
    vector[Tp_card[p]] Theta;
    array[Tp_card[p]] int Subj_p = segment(Subj, pos, Tp_card[p]);
    array[Tp_card[p]] int S_p = segment(S, pos, Tp_card[p]);
    sx = (p-1)*Q+1;
    ex = p*Q;
    
    // Calculate
    Phi_mat = B * Psi[sx:ex,]; 
    Mu = B * w_mu[sx:ex];
    Theta = rows_dot_product(Scores[Subj_p, ], Phi_mat[S_p, ]);

    // Likelihood and increment
    segment(Y, pos, Tp_card[p]) (*$\sim$*) normal(Mu[S_p]+Theta, sqrt(sigma2[p]));
    pos = pos + Tp_card[p];
  }
}
\end{lstlisting}
\doublespacing
\section{Simulations}\label{sec:MSFAST_sim}

To evaluate MSFAST, we adapt the simulation design from \cite{happ_multivariate_2018}. All simulation scripts are available on \href{https://github.com/JSartini/MSFAST_B-m-FPCA}{GitHub}. Data are simulated from Model~\eqref{EQ:Multivariate_FPCA} setting
\begin{itemize}
\centering
    \item[] $\mu^{(p)}(t) = (-1)^p\cdot 2\sin((2\pi + p)t)$, $t \in [0,1]$,
    \item[] $\phi^{(p)}_k(t) = \begin{cases}(-1)^{p}\times \sqrt{\frac{2}{P}}\sin[2 (\lfloor \frac{k-1}{2}\rfloor + 1)\pi t] \text{, $k$ odd}\\
    (-1)^{p}\times \sqrt{\frac{2}{P}}\cos[2(\lfloor \frac{k-1}{2}\rfloor + 1)\pi t]\text{, $k$ even} \end{cases}$, 
\end{itemize}
with eigenvalues $\lambda_k = 0.5^{k-1}$ for $k=1,2,3$, and signal-to-noise ratios ${\rm SNR}_p = \sum_{k=1}^3\lambda_k/\sigma^2_{p}$. Here, $P$ = 3 and ${\rm SNR}_p$ is set to $4$ for all covariates. Data are initially sampled on an equally spaced grid of size $M = 100$ for $I = 100$ subjects. For each subject $i$ and function $p$, we sample the number of observations $J_i^{(p)}$ either: (1) uniformly between $3$-$7$ ($5$ expected observations);  or (2) uniformly between $5$-$15$ ($10$ expected observations). We then sample $J_i^{(p)}$ observations without replacement among the $M=100$ equally spaced potential observations separately for each covariate; all other observations are left missing. The resultant observation times can vary within subject across covariates. For all scenarios, we use $B = 500$ simulations and spline basis dimension $Q = 20$.

MSFAST is compared to: (1) a univariate sparse FPCA approach which does not model correlations between variates (``uFPCA'') \citep{xiao_fast_2018}; (2) variational Bayes using message passing (``VMP''); (3) variational Bayes using mean field approximation (``MF'') \citep{nolan_efficient_2025}; (4) \textit{Fast Covariance Estimation for Multivariate Sparse Functional Data} (``mFACES'') \citep{li_fast_2020}; and (5) \textit{Multivariate FPCA for Data Observed on Different (Dimensional) Domains} (``mFPCA'') \citep{happ_multivariate_2018}.

The first evaluation criterion is estimation accuracy quantified by relative integrated squared error (RISE). Integrated squared error for variate $p$ (denoted ${\rm ISE}^{(p)}$) is defined as:
\begin{equation}\label{EQ:Metrics_ISE}
\begin{split}
        {\rm ISE}^{(p)} & = \frac{1}{I}\sum_{i = 1}^I\int_0^1\{\widehat{Y}^{(p)}_i(t) - Y_{i, {\rm true}}^{(p)}(t)\}^2dt \\
        & \approx \frac{1}{I} \sum_{i = 1}^I \sum_{m=1}^M q(t_m)\{\widehat{Y}_i^{(p)}(t_m) - Y^{(p)}_{i, {\rm true}}(t_m)\}^2\;,
\end{split}
\end{equation}
where $q(\cdot)$ are quadrature weights and $\widehat{Y}^{(p)}_i(\cdot)$ is the estimator of the unknown true $Y_{i, {\rm true}}^{(p)}(\cdot)$ for a simulation and method. Here we focus on relative ISE (RISE): $\text{RISE}_{\text{Method}}^{(p)}=\text{ISE}_{\text{Method}}^{(p)}/\text{ISE}_{\text{Mean}}^{(p)}$, where $\text{ISE}_{\text{Mean}}^{(p)}$ is ${\rm ISE}^{(p)}$ when the subject-specific trajectories are estimated by the corresponding sample means ($\widehat{Y}_i^{(p)}(t) = \frac{1}{J_i^{(p)}}\sum_{t \in \mathbf{T}_i^{(p)}} Y_i^{(p)}(t)$).

The second criterion is the coverage probability of pointwise $95$\% confidence/credible intervals for $Y^{(p)}_{i, {\rm true}}(t)$. For every simulation $b = 1, \ldots, B$ and time point along the potential observation grid, we obtain an equal-tail $95$\% confidence/credible interval and record whether the true observation is covered. Aggregating these indicators over subjects $i$ and potential observation points $t$ provides an estimate of coverage for each method, simulation $b$, and variable $p$.

\begin{figure}[h]
\centering\includegraphics[width=15cm]{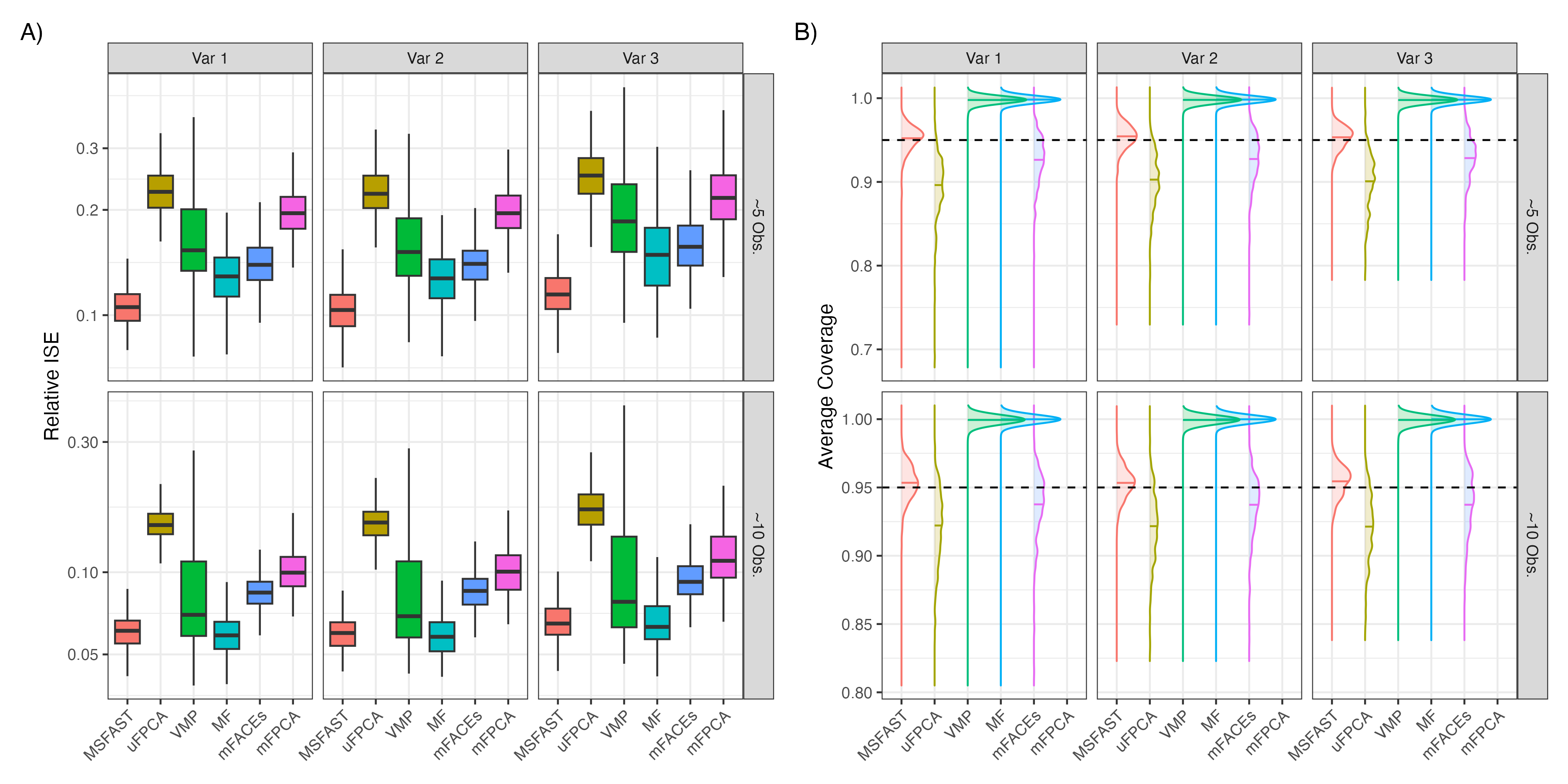}
\caption{\textbf{A)} Boxplots of RISE and \textbf{B)} kernel smoother of 95\% interval coverage probabilities of the underlying smooth functions for MSFAST, uFPCA, VMP, MF, mFACEs, and mFPCA. Columns correspond to covariate and rows to expected number of observations ($5$ then $10$).}
\label{fig:Smooths_Perf_MV}
\end{figure}

Figure~\ref{fig:Smooths_Perf_MV} displays the boxplots of RISE in Panel \textbf{A} and kernel density plots of average pointwise coverage for $95\%$ confidence/credible intervals for $Y^{(p)}_{i, {\rm true}}(t)$ in Panel \textbf{B}. Within each sub-panel, columns correspond to functional variate and rows correspond to average number of observations per subject, $5$ in the first row and $10$ in the second. The x-axis labels and colors indicate the method. In Panel \textbf{A}, all approaches reduce ISE relative to taking the mean, but MSFAST has the lowest RISE distribution. From Panel \textbf{B}, MSFAST has the closest to nominal mean coverage. The variational approaches are conservative, uFPCA and mFACEs produces sub-nominal coverage, and mFPCA does not produce confidence intervals for the true smooth functions.

We also compared the accuracy of $\mu^{(p)}(t)$ and $\phi^{(p)}_k(t)$ estimates, quantified by ISE. The ``uFPCA'' method was excluded, as it does not estimate the multivariate model. For each simulation and functional component $f(\cdot)$, we calculate:
\begin{equation}
    \begin{split}
        {\rm ISE}_f & = \int_0^1 \{\widehat{f}(t) - f^{true}(t)\}^2dt \approx \sum_{m=1}^M q(t_m)\{\widehat{f}(t_m) - f^{\rm true}(t_m)\}^2\;,
    \end{split}
    \label{EQ:Metrics_ISE_funcs}
\end{equation}
where $q(\cdot)$ are quadrature weights and $\widehat{f}(\cdot)$ is the estimate of $f^{\rm true}(\cdot)$ for the chosen iteration and method. The evaluation time points $t_m$ again come from the potential observation grid. We apply the Procrustes postprocessing alignment described in Section~\ref{sec:ACE} to all methods, using $\widetilde{\boldsymbol{\Phi}} = \boldsymbol{\Phi}^{\rm true}$, before calculating ISE.

\begin{figure}[h]
\centering\includegraphics[width=12cm]{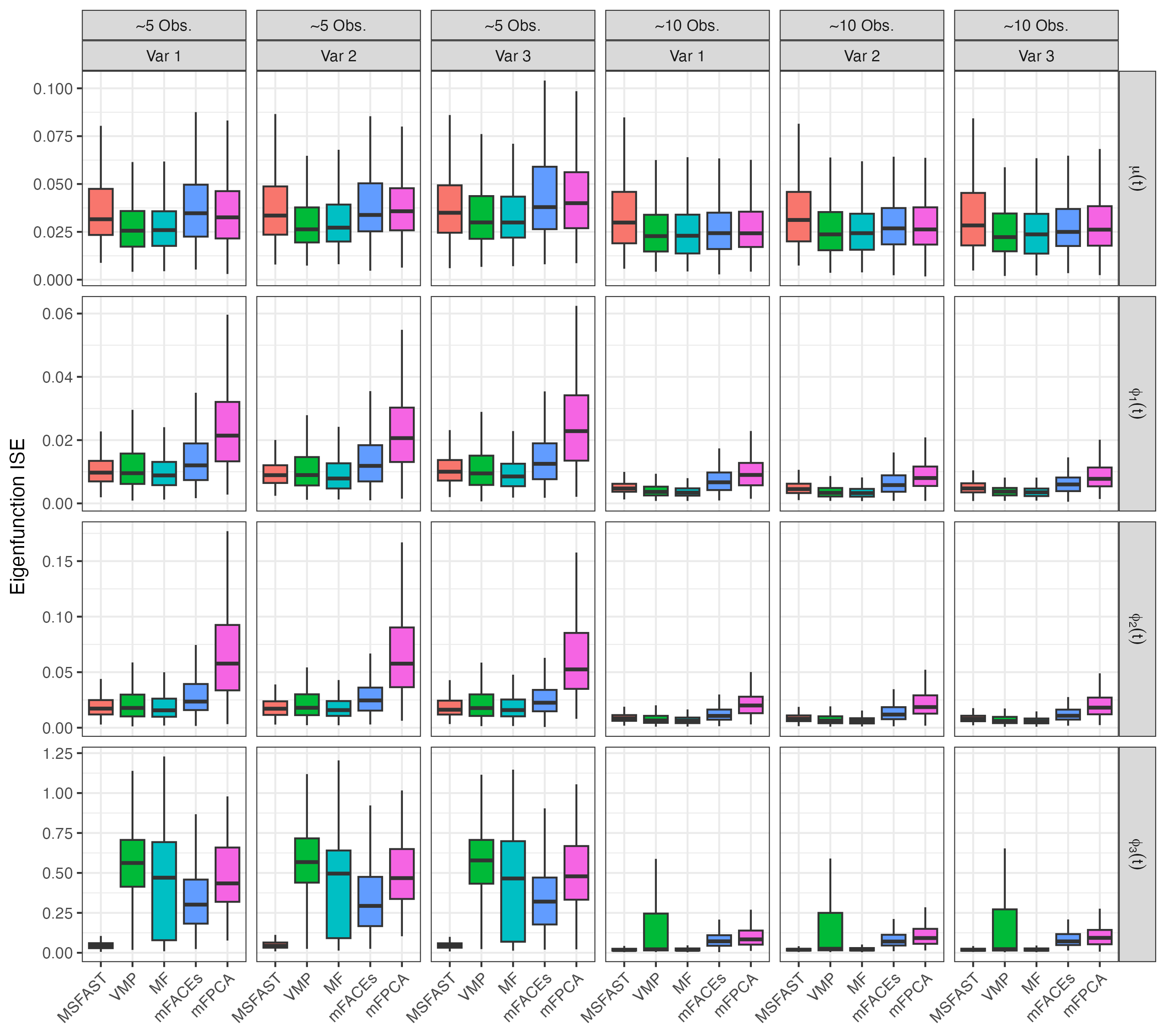}
\caption{Mean and eigenfunction ISE for MSFAST, VMP, MF, mFACEs, and mFPCA. Columns $1$-$3$: $5$ expected observations; columns $4$-$6$: $10$ expected observations. Columns $1$, $4$: variable 1; columns $2$, $5$: variable 2; columns $3$, $6$: variable 3. Rows correspond to the means, $\mu^{(p)}(t)$, and first three FPCs, $\phi^{(p)}_k(t)$.}
\label{fig:Funcs_ISE_MV}
\end{figure}

Figure~\ref{fig:Funcs_ISE_MV} displays the ISE values for estimating the mean and eigenfunctions components corresponding to each functional variable. MSFAST produces ISE values comparable or superior to the other methods, with the largest differences when data are more sparse and the signal is smaller (eigenfunctions with smaller eigenvalues). MSFAST also produces nearly nominal mean coverage of these components; for more details, see Supplement Section~\ref{supp:sims}. The comparators do not produce confidence bounds for the mean or FPCs. While mFPCA can produce bootstrap confidence intervals of the FPCs, this returns errors traced to the estimation of univariate FPCs. Adjusting the univariate FPCA implementation may solve this problem, but we did not change existing implementations.

We finally evaluated the computational performance of MSFAST relative to the comparators. Using the same simulation scenario with $K = 3$, we fixed $5$ expected observations per subject, $M = 2000$ potential observation points, and $P = 3$ variates, scaling the number of subjects $I \in \{100, 200, 300, 400, 500, 1000\}$. We then fixed $I = 100$ and varied the number of variates $P \in \{3, 4, 5, 6, 7, 8\}$. We fit all methods to 5 simulated datasets on a personal laptop (MacBook Pro, 3.49 GHz, 32GB RAM), recording the computational time taken. MSFAST was not parallelized for this comparison. 

\begin{figure}[h]
\centering\includegraphics[width=12cm]{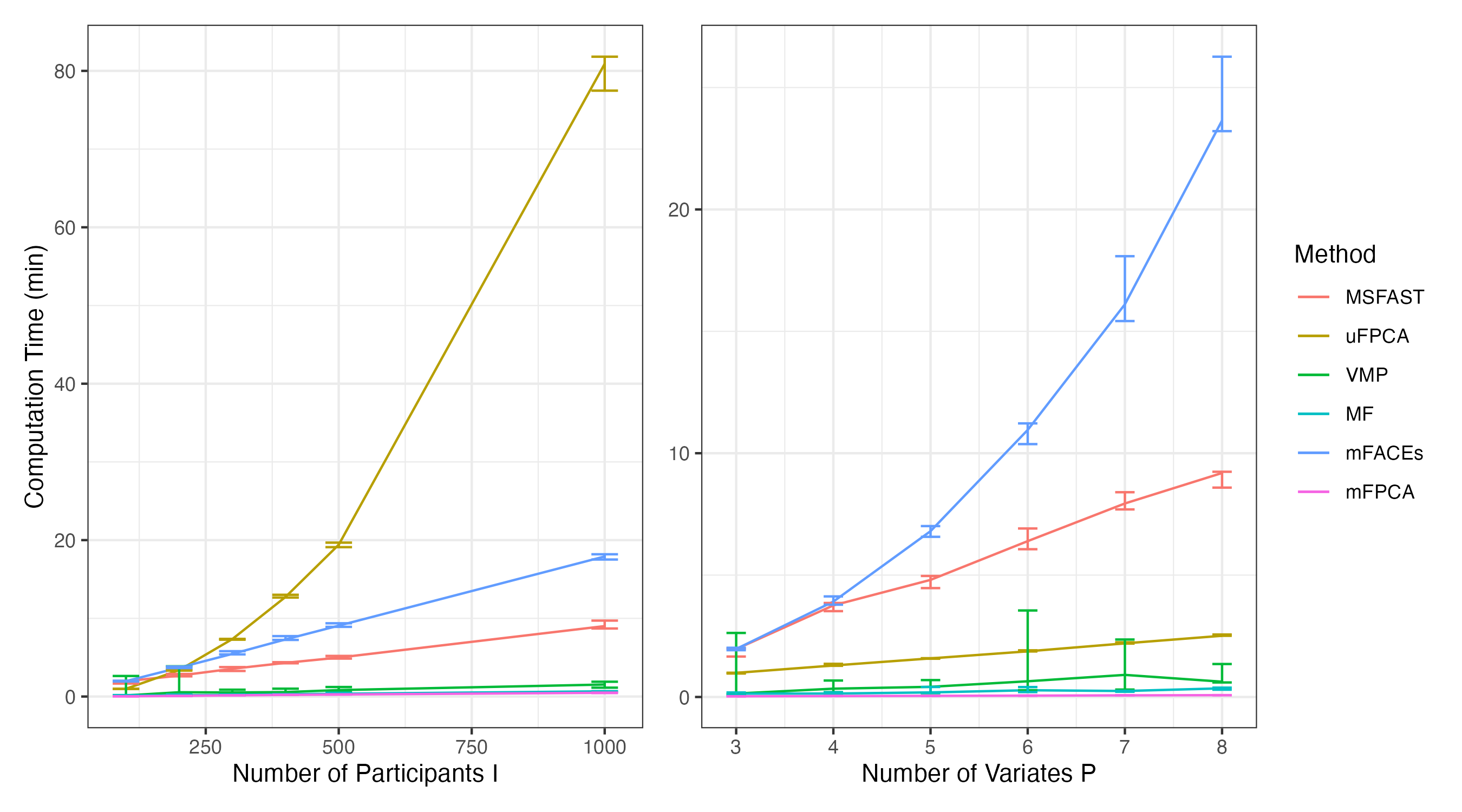}
\caption{\textbf{A)} Computation time (y-axis in minutes) as a function of number of subjects $I$ (x-axis) and \textbf{B)} as a function of the number of variates $P$ for MSFAST, uFPCA, VMP, MF, mFACEs, mFPCA.
Lines: median time; error bars: min and maximum time.}
\label{fig:Timing_MV}
\end{figure}

Figure~\ref{fig:Timing_MV} displays the median (line plot) and extrema (error bars) of computation times for each method. All methods other than uFPCA scale approximately linearly in the number of subjects $I$, while all methods other than mFACEs appear to scale linearly in the number of covariates $P$. MSFAST is not faster than the variational methods or mFPCA, but it takes less time than mFACEs as both $I$ and $P$ scale. Moreover, MSFAST computation time remains feasible, (less than $10$ minutes for $I = 1000$ and $P = 8$. Computation times of mFACEs and uFPCA are likely impacted by steps taken to mitigate memory constraints during prediction.

It is important to note that these methods have different computational bottlenecks. Estimation of the smoothed covariance theoretically dominates mFPCA and mFACEs, so their computation times should scale with the number of unique domain points observed \citep{happ_multivariate_2018, li_fast_2020}. MSFAST, VMP, and MF instead scale with the number of likelihood evaluations, regardless of observation times. This means that mFPCA and mFACEs should scale sub-linearly when there is a constraint on the number of unique observation points (see Supplement Section~\ref{supp:sims}). This implies that, for large studies with strong signals, mFPCA and mFACES may be computationally preferable to the Bayesian approaches.

\section{Application: the CONTENT Study}\label{sec:applications}

MSFAST was applied to the CONTENT child growth study, where multiple growth measures are taken at the same time for each participant but sparsely across subjects. The CONTENT child growth study was conducted between May 2007 and February 2011 in Las Pampas de San Juan Miraflores and Nuevo Paraíso, two peri-urban shanty towns located on the southern edge of Lima City in Peru. The towns had approximately $40{,}000$ residents with $25$\% of the population under the age of $5$ \citep{checkley_effects_1998,checkley_effects_2003}. A simple census was conducted to identify pregnant women and children less than $3$ months old. Eligible newborns and pregnant women were randomly selected and invited to participate in the study (at most one newborn per household). This cohort study aimed to assess whether Helicobacter pylori (H. pylori) infection adversely affects the growth in children less than $2$ years of age \citep{jaganath_first_2014, crainiceanu2024book}. The study collected length and weight measures weekly until the child was $3$ months old, biweekly between $3$ and $11$ months, and once monthly afterwards. Missed and canceled visits contributed to the sparse data structure. We focus on the length and weight z-scores relative to the age- and sex-specific World Health Organization (WHO) standards. In this paper, we analyze a subset of these data ($n = 197$ children) publicly available as part of the {\ttfamily refund} package in R \citep{goldsmith_refund_2010}; see also the website \href{http://www.Functionaldataanalysis.org}{FDAwR} accompanying the  monograph Functional Data Analysis with R \citep{crainiceanu2024book}. An accompanying vignette, available on \href{https://github.com/JSartini/MSFAST_B-m-FPCA}{GitHub}, details this analysis step-by-step.

\begin{figure}[h]
\centering\includegraphics[width=12cm]{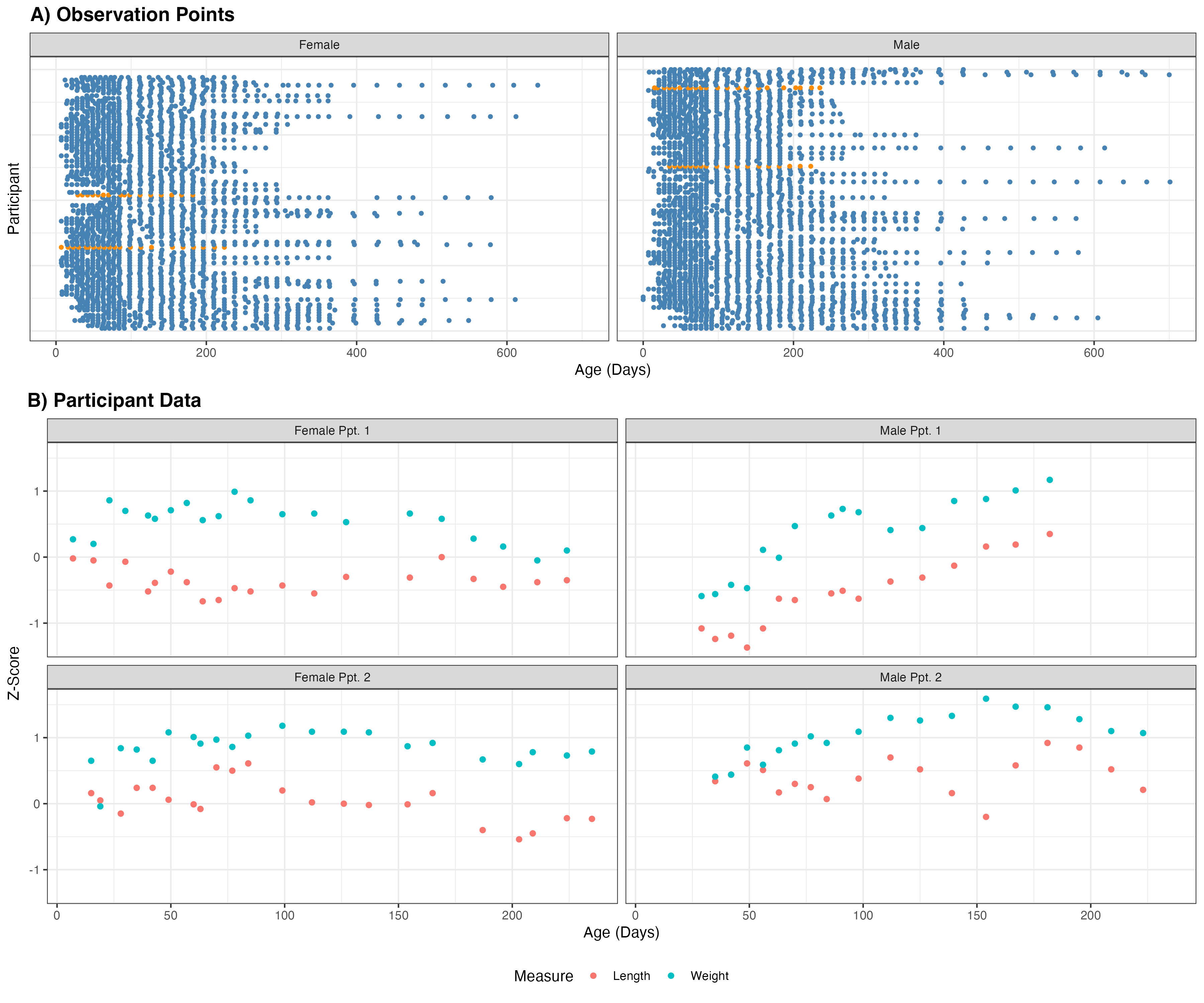}
\caption{\textbf{A)} Length and weight observation times for each CONTENT participant stratified by sex. Y-axis: participant; x-axis: age in days. Highlighted points correspond to the data in Panel \textbf{B}. \textbf{B)} Data from two randomly chosen participants within each gender, color-coded by measure. Y-axis: z-score; x-axis: age in days.}
\label{fig:Content_Obs}
\end{figure}

Figure~\ref{fig:Content_Obs} is a visualization of the CONTENT data structure. Panel \textbf{A} shows the sampling times (each child on a horizontal line) stratified by gender (females in the left and males in the right column). Four individuals, two females and two males, are highlighted in Panel \textbf{A}. Panel \textbf{B} provides length and weight z-scores relative to the age and sex-based WHO standards for these four children.

We used MSFAST for joint modeling and prediction of trajectories at points when data were not observed, withholding two randomly chosen subjects for dynamic prediction (see Sections~\ref{sec:Multivariate} and \ref{sec:prediction}). The number of FPCs was set to $K=4$, as our sensitivity analyses indicated this explained $95\%$ of variability in the data (Supplement Section~\ref{supp:var_exp}). The dimension of $\mathbf{B}(t)$ was set to $Q=20$ to provide a sufficiently rich basis. We produced a posterior FPC estimate using the routine described in Section~\ref{sec:ACE}, then aligned all FPC samples with this estimate using the Procrustes transform described in the same Section. MSFAST was run for 3000 iterations, discarding the first 2000 as burn-in, taking $\approx 12$ minutes to run on the personal laptop used in Section~\ref{sec:MSFAST_sim}, and all R-hat statistics after alignment were less than $1.01$, indicating convergence. 

\begin{figure}[h]
\centering\includegraphics[width=12cm]{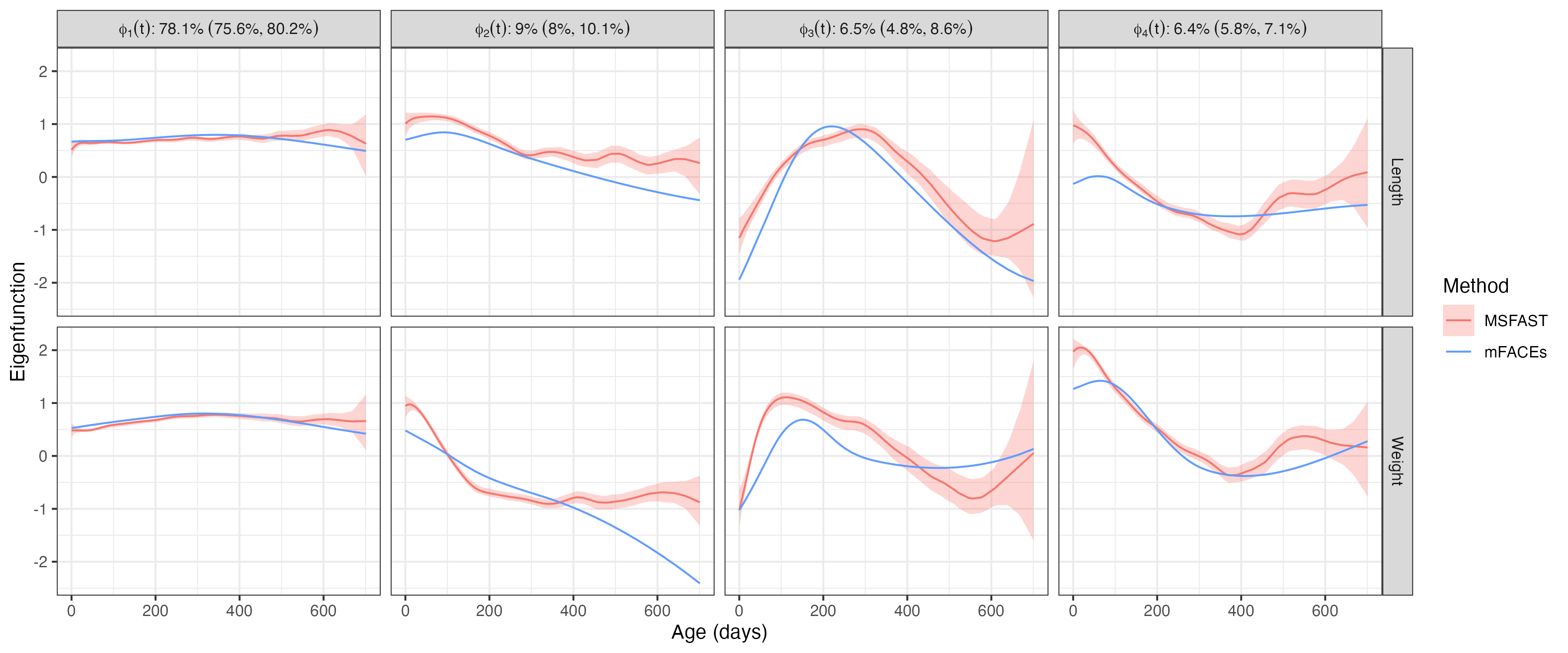}
\caption{Comparison between MSFAST and mFACEs of posterior inference on the FPCs of the Content data. Panel rows indicate measure, while columns correspond to FPC. Facet labels indicate percent variance explained among the 4 FPCs with credible interval. X-axis: age in days.}
\label{fig:Content_Decomp}
\end{figure}

Figure~\ref{fig:Content_Decomp} displays the resulting FPC estimates and credible intervals obtained from MSFAST (red) and corresponding mFACEs estimates (blue). As indicated in Section~\ref{sec:MSFAST_sim}, mFACEs does not perform uncertainty quantification on the FPCs, so no confidence intervals can be shown. Panel rows indicate growth measure, length then weight. Notably, the widths of MSFAST credible intervals increase as a function of age, reflecting the greater sparsity of the data and lesser information beyond one year of age; see Figure~\ref{fig:Content_Obs}. Figure~\ref{fig:Content_Decomp} also indicates that the first FPC (an intercept with minimal curvature) explains $78.1$\% of variance among the FPCs. Finally, MSFAST exhibits similar smoothness and FPC shape to mFACEs. 

\begin{figure}[h]
\centering\includegraphics[width=12cm]{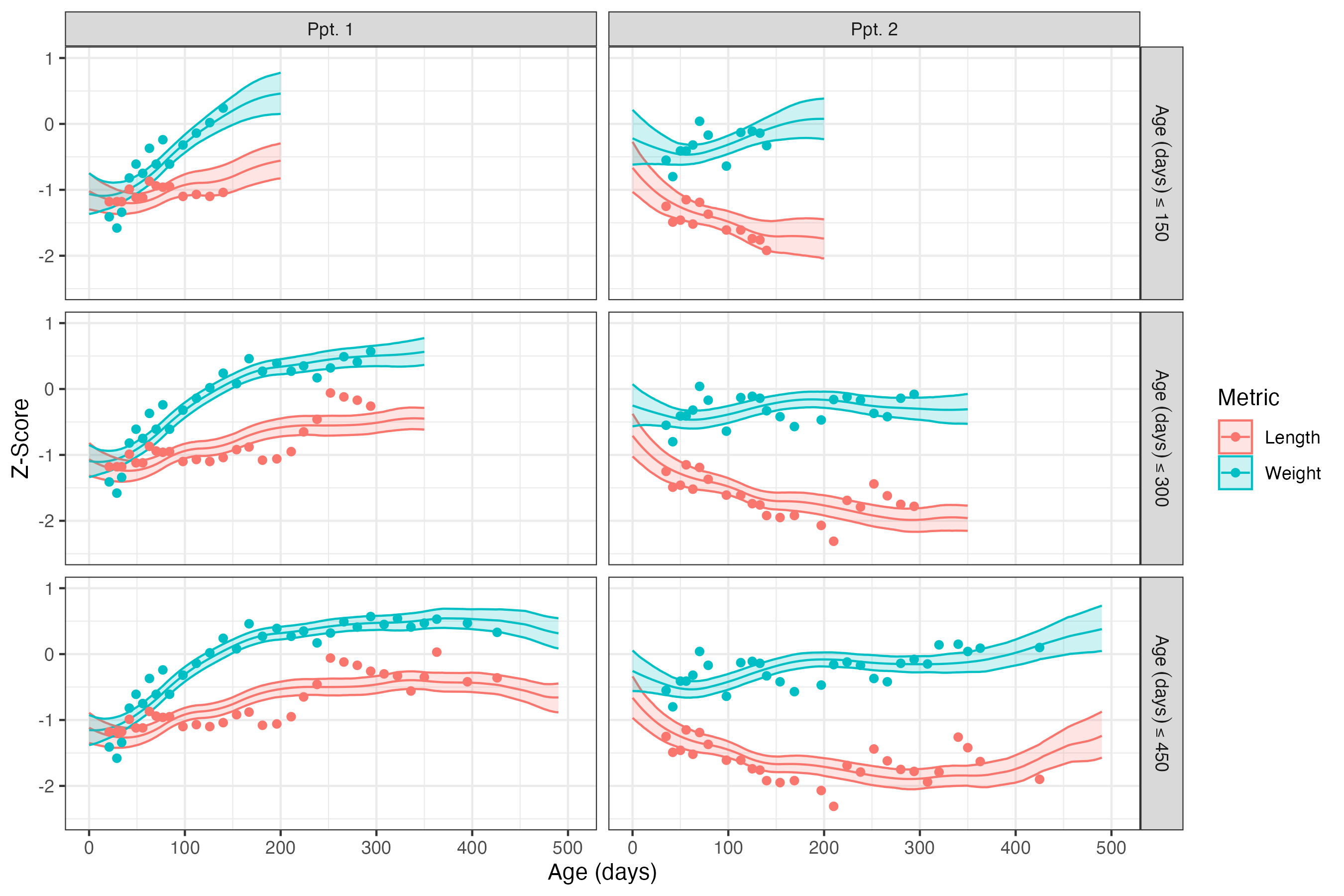}
\caption{Observed data (points), dynamic predictions from MSFAST (solid lines), and $95$\% pointwise credible intervals (shaded areas) of age and sex adjusted z-scores of length and weight. Panel columns: CONTENT participants; panel rows: data window used in prediction. Y-axis: z-score scale; x-axis: age in days.}
\label{fig:Content_Pred}
\end{figure}

Dynamic prediction is conducted for the two withheld children using the score resampling method described in Section~\ref{sec:prediction}. We form predictions using data up to $150, 300,$ and $450$ days of age, restricting prediction to $50$ days ahead of the observed window. We visualize these predictions, updating them as more data become available, in Figure~\ref{fig:Content_Pred}. Each panel corresponds to a particular individual (column) and prediction (row). The observed data used in prediction is indicated by points, the predicted trajectories are shown using solid lines, and the $95$\% credible intervals for those trajectories are indicated by shaded areas around these lines. 

Figure~\ref{fig:Content_Pred} indicates that the dynamic predictions of latent trajectories track the data well and provide reasonably smooth estimates. Indeed, it is hard to believe that the large week-to-week fluctuations in observed z-scores are real, whereas the smooth predicted curves are more consistent with what is known about human biology. The width of the credible intervals is smaller where growth data are observed. As data becomes sparser, the credible intervals begin to widen. However, these predictions leverage information from study participants who have data at these time points, resulting in much tighter prediction intervals than if there were no information borrowing across study participants. 

Let us investigate more closely the data for the first child (left panels in Figure~\ref{fig:Content_Pred}). This child has a sharp increase in their weight z-score from $\approx -1$ after birth to $\approx 0$ by day $150$. The prediction of future weight at day $200$ when data are available up to day $150$ (top first row panel) indicates a moderation in this growth rate, likely due to the slower length z-score growth (red) and the known growth patterns of other children in this study. This is exactly what happens in the later panels, with the length stabilizing around $-0.5$ and weight stabilizing around $0.5$. 

\section{Discussion}\label{sec:MSFAST_disc}
MSFAST provides a fully-Bayesian approach to multivariate, sparse data FPCA, a useful tool for analyzing emerging multi-modal data with complex sampling structure. MSFAST accounts for the variability associated with estimating the PCs while remaining computationally feasible. As an extension of FAST, MSFAST leverages projection of the FPCA basis onto a rich orthonormal spline basis and polar decomposition to efficiently sample the orthonormal matrix of spline basis coefficients. MSFAST addresses the more challenging scenario of multivariate, sparse data through pre- and post-processing routines and precise adjustments to the Bayesian MCMC modeling structure. Using this specific combination of modeling techniques, MSFAST can be implemented in any Bayesian modeling software, such as {\ttfamily STAN}.

Input data for MSFAST is first standardized within covariate. This ensures that covariates of disparate scales cannot degrade the posterior geometry conditioning due to numerical stability issues or improper smoothing of variates with smaller scale. Input data is further ordered by variable, allowing for the likelihood calculations to be parallelized. 

MSFAST uses a different orthogonal spline basis, one which no longer aims to represent signals over contiguous sub-intervals of the functional domain using a small subset of the splines. This prevents gaps in the observed data from leading to model non-identifiability. The Bayesian model underpinning MSFAST is also adjusted. The FPC scores now use a non-centralized parameterization to improve sampling stability.

After MSFAST is run, outputs are re-scaled to align with the original data, and the FPCs can be reliably aligned using a post-processing routine based upon the Procrustes transformation. This routine ensures that variation between FPC samples reflects only changes in the FPC basis span. If dynamic prediction of newly observed subjects is desired after fitting, the model formulation of MSFAST facilitates efficiently sampling the FPC scores for the newly observed subject conditional on the population-level parameter samples and the observed data. 

There are still multiple open questions, including: (1) whether these results generalize to non-Gaussian outcome data; (2) how to implement non-uniform priors on the Stiefel manifold to further reduce multi-modality (e.g., eliminate the sign-flipping inherent to PCA); (3) how to best estimate the dimension of the FPC basis ($K$); and (4) theoretical guarantees on the convergence of the MSFAST posterior estimates.

\section{Software}
\label{sec5}

Software in the form of R and STAN code is available on
\href{https://github.com/JSartini/MSFAST_B-m-FPCA}{GitHub}.

\section{Data availability}

The CONTENT data are available as part of the {\ttfamily refund} R package, and can be accessed by installing the package and calling {\ttfamily data(content)}.

\clearpage
\appendix
\section*{\centering SUPPLEMENTARY MATERIAL}
\setcounter{figure}{0}
\setcounter{table}{0}
\setcounter{section}{0}
\setcounter{equation}{0}
\renewcommand{\thefigure}{S\arabic{figure}}
\renewcommand{\thetable}{S\arabic{table}}
\renewcommand{\thesection}{S\arabic{section}}
\renewcommand{\theequation}{S\arabic{equation}}
\onehalfspacing

\subsection*{Penalty Matrix $\mathbf{P}$}\label{supp:penalties}

We chose a spline smoothing prior based on the roughness penalty $\alpha \int f^2(t) dt + (1-\alpha) \int \{f''(t)\}^2 dt$. For spline parameters $\theta$ such that $f(t) = \mathbf{B}(t)\theta$, there exist unique penalty matrices $\mathbf{P}_0, \mathbf{P}_2$ such that $\int f^2(t) dt= \theta^\top \mathbf{P}_0\theta$ and $\int \{f''(t)\}^2 dt = \theta^\top\mathbf{P}_2\theta$ \citep{kimeldorfwahba1970,cravenwahba1979,Wahba1983,osullivan1986}. We follow \cite{goldsmith_generalized_2015} and define $\mathbf{P}_{\alpha} = \alpha \mathbf{P}_0 + (1-\alpha) \mathbf{P}_2$. This allows us to write the final penalty using a single quadratic form: 
\begin{align*}
    \alpha \int f^2(t) dt + (1-\alpha)\int \{f''(t)\}^2 dt & = \alpha \theta^\top \mathbf{P}_0 \theta + (1- \alpha) \theta^\top \mathbf{P}_2 \theta \\
    & = \theta^\top(\alpha \mathbf{P}_0 + (1-\alpha) \mathbf{P}_2) \theta\\
    & = \theta^\top\mathbf{P}_{\alpha}\theta
\end{align*}
We proceed by defining both of the penalty components $\mathbf{P}_0, \mathbf{P}_2$ separately, as $\mathbf{P}_{\alpha}$ is just a linear combination of these matrices. Let $\mathbf{B}(t) = [B_1(t)|\ldots,|B_Q(t)]$ represent the basis functions, chosen to be orthonormal in $L_2[0,1]$. First, we define the zero-order penalty $\mathbf{P}_0$ element-wise. This derivation leverages the orthonormal definition of the $B_i(t)$.
\begin{align*}
    (\mathbf{P}_0)_{ij} & = \int_0^1 \int_0^1 B_i(t) B_j(t) dt\\
    & = \begin{cases} 1 \text{ when } i = j\\
    0 \text{ otherwise} \end{cases}
\end{align*}
The resulting matrix $\mathbf{P}_0 = \mathbf{I}_Q$ by the definition of the basis $\mathbf{B}(t)$. This clearly indicates the role of this component (when non-zero) in ensuring the non-singularity of the final penalty $\mathbf{P}_{\alpha}$, similar to adding a ridge penalty in the context of regression.

Next, we define the more central ``wiggliness" penalty $\mathbf{P}_2$. For this penalty, based on the squared second derivative, we introduce the second derivatives of the basis $\mathbf{B}(T)$: $\mathbf{B}''(t) = [B_1''(t)|\ldots,|B_Q''(t)]$. We are able to quickly retrieve these derivatives using the properties of B-splines, from which the orthogonalized B-splines are derived \citep{redd_comment_2012}. The elements of $\mathbf{P}_2$ are as follows.
\begin{align*}
    (\mathbf{P}_2)_{ij} & = \int_0^1 \int_0^1 B''_i(t) B_j''(t) dt
\end{align*}

Combining $\mathbf{P}_0$ and $\mathbf{P}_2$ in linear combination produces the penalty $\mathbf{P}_{\alpha}$, which will be non-degenerate when $\alpha > 0$ (the weight of the absolute penalty is non-zero).

\subsection*{Sensitivity Analysis of Spline Basis Dimension $Q$}\label{supp:q_sens}

To evaluate the dependence of MSFAST on the chosen spline basis dimension $Q$, we leveraged the simulation structure in Section~\ref{sec:MSFAST_sim}, fixing $P = 3$, $K = 3$, $I = 100$, $M = 100$, penalty hyper-parameter $\alpha = 0.1$, and signal-to-noise ratio of $4$, all while varying $Q$ between 5 and 40. This provides some idea of the effects of moving from a very restrictive to a very rich basis. We evaluated the ability of MSFAST to capture the latent smooth functions under each $Q$ value using the estimation and inference metrics introduced in Section~\ref{sec:MSFAST_sim}.

\begin{figure}[h]
\centering
\includegraphics[width=12cm]{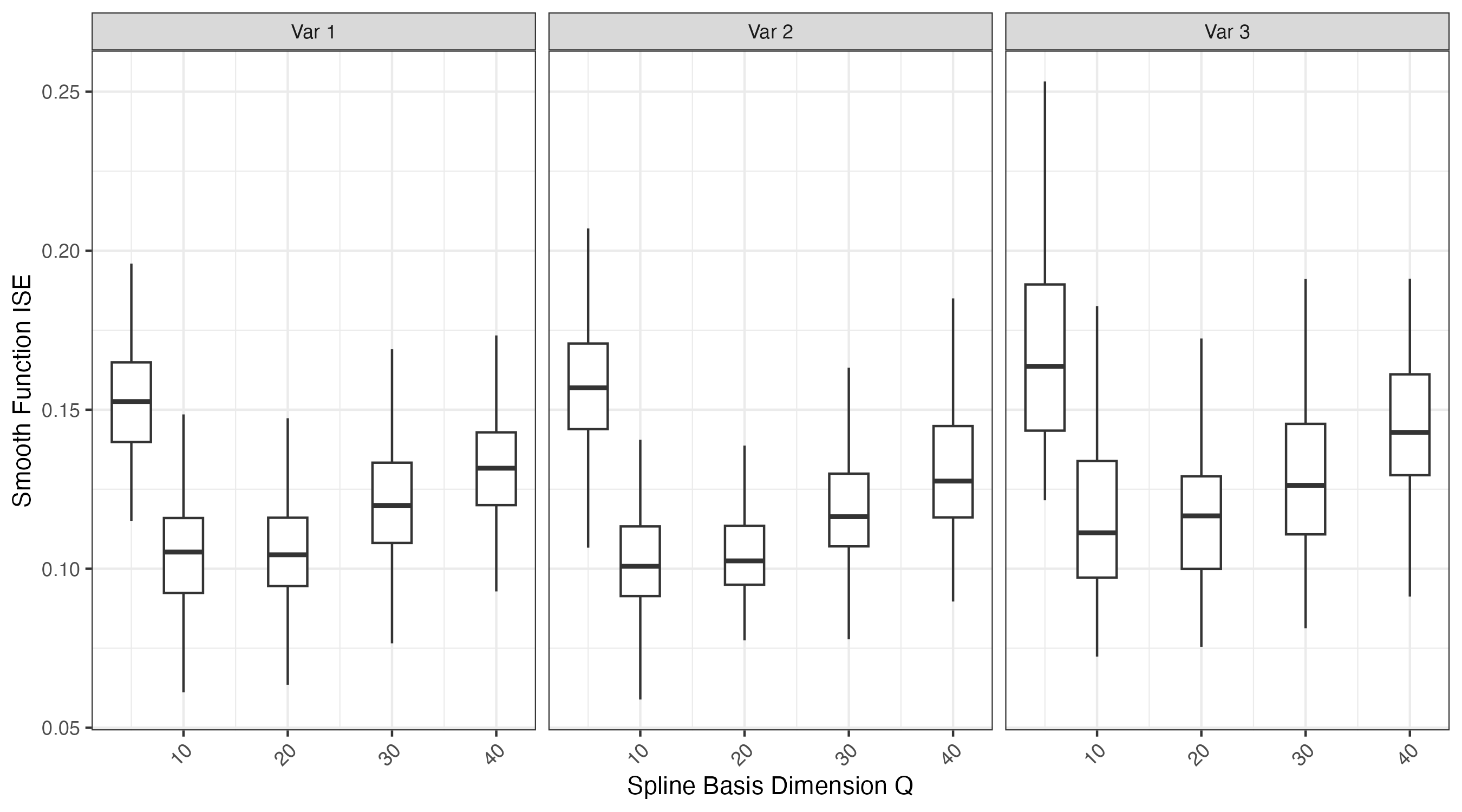}
\caption{Boxplots of RISE from applying MSFAST using different spline basis dimensions $Q$. Panels correspond to variate.}
\label{supp_fig:Q_ISE}
\end{figure}

Figure~\ref{supp_fig:Q_ISE} indicates setting $Q = 5$ produces estimates with greater error, though they are still superior to taking the simple mean. This is likely due to the lack of flexibility in the basis. However, the RISE quickly stabilizes around spline basis size $Q = 10\text{-}20$.

\begin{figure}[h]
\centering
\includegraphics[width=12cm]{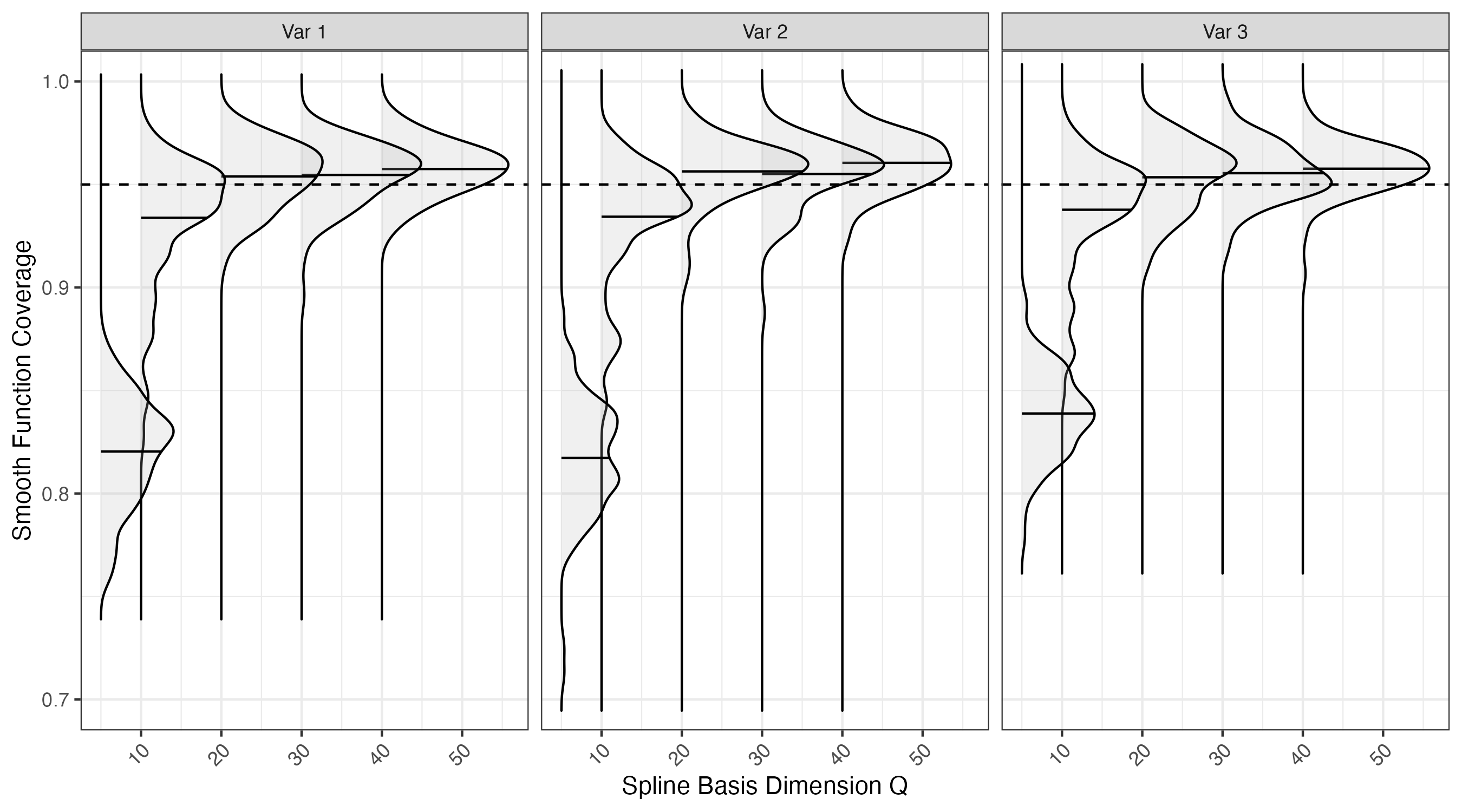}
\caption{Kernel smooths of coverage for the $95\%$ credible intervals of latent smooth functions produced by MSFAST under different spline basis dimensions $Q$. Panels correspond to variate.}
\label{supp_fig:Q_COV}
\end{figure}

Figure~\ref{supp_fig:Q_COV} similarly shows that all values $Q \geq 20$ produce nominal, stable inference.

\subsection*{Proof of Prior Propriety}\label{supp:proper_prior}

First, we provide the form $p(\boldsymbol{\psi}_k|h_k)$ of the priors for the eigenfunction spline weight vectors $\boldsymbol{\psi}_k \in \mathbb{R}^{PQ}$, conditional on their corresponding smoothing parameters $h_k$ up to a constant. Here, we let $\mathbf{\Psi} = [\boldsymbol{\psi}_1| \ldots | \boldsymbol{\psi}_K]$ and use $R$ to represent the rank of the penalty matrix $\mathbf{P}_\alpha$ as detailed in Supplement Section~\ref{supp:penalties}.
\begin{equation}\label{eqn:MSFAST_condprior}
p(\boldsymbol{\psi}_k|h_k) = h_k^{PR/2} \exp(-h_k\boldsymbol{\psi}_k^\top (\mathbf{I}_p \otimes \mathbf{P}_\alpha) \boldsymbol{\psi}_k/2) \times \mathbbm{1}(\boldsymbol{\Psi} \in \mathcal{V}_{K, PQ})
\end{equation}
We can combine the conditional prior in Equation~\ref{eqn:MSFAST_condprior} with the independent gamma priors on the $h_k$, which we parameterize here as $\Gamma(h_k|\alpha_\psi, \beta_\psi)$ for shape $\alpha_\psi$ and rate $\beta_\psi$, to produce the joint prior on the eigenfunctions and their smoothing parameters in Equation~\ref{eqn:full_prior} (up to a normalizing constant). We denote this prior as $p(\boldsymbol{\Psi}, \mathbf{H})$ for the set of smoothing coefficients $\mathbf{H} = \{h_1, \ldots, h_K\}$.
\begin{equation}\label{eqn:full_prior}
\begin{split}
    p(\boldsymbol{\Psi}, \mathbf{H}) & = \prod_{k = 1}^K h_k^{PR/2} \exp\{-h_k\boldsymbol{\psi}_k^\top (\mathbf{I}_p \otimes \mathbf{P}_\alpha) \boldsymbol{\psi}_k/2\} \Gamma(h_k|\alpha_\psi, \beta_\psi) \times \mathbbm{1}(\boldsymbol{\Psi} \in \mathcal{V}_{K, PQ})\\
    & = \exp\{-\frac{1}{2}\sum_{k = 1}^K h_k \boldsymbol{\psi}_k^\top (\mathbf{I}_P \otimes \mathbf{P}_\alpha)\boldsymbol{\psi}_k/2\} \prod_{k = 1}^K h_k^{PR/2}\Gamma(h_k|\alpha_\psi, \beta_\psi) \times \mathbbm{1}(\boldsymbol{\Psi} \in \mathcal{V}_{K, PQ})
\end{split}
\end{equation}
This has the same form as the joint prior addressed in \cite{sartini_2026_suffcond}, so we can apply the main result from that chapter to find that this prior (and thus the posterior) is finite for any valid choice of hyper-parameters $\alpha_\psi > 0$ and $\beta_\psi > 0$. This concludes proof of Result~\ref{rmk:proper_prior}.

\subsection*{Sensitivity Analysis of Penalty Parameter $\alpha$}\label{supp:alpha_sens}

To evaluate the dependence of MSFAST on the chosen penalty hyper-parameter $\alpha$ (which is described in Supplement Section~\ref{supp:penalties}), we leveraged the simulation structure in Section~\ref{sec:MSFAST_sim}, fixing $P = 3$, $K = 3$, $Q = 20$, $I = 100$, $M = 100$, and signal-to-noise ratio of $4$, all while varying $\alpha$ between $0.01$ and $0.3$. This provides some idea of the effects of varying the degree of zeroth order penalization within the smoothing procedure. We evaluated the ability of MSFAST to capture the latent smooth functions under each $\alpha$ value using the estimation and inference metrics introduced in Section~\ref{sec:MSFAST_sim}.

\begin{figure}[h]
\centering
\includegraphics[width=12cm]{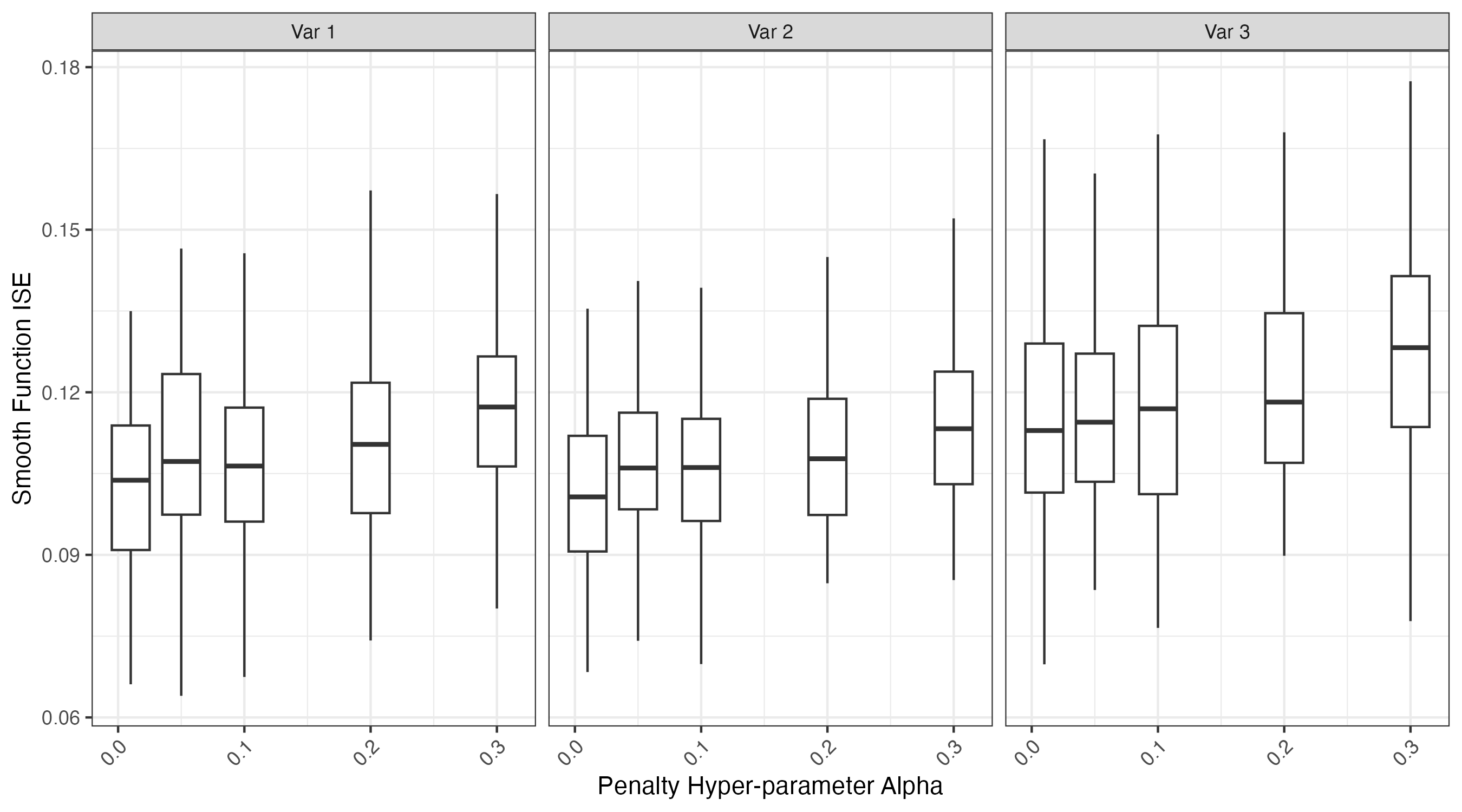}
\caption{Boxplots of RISE from applying MSFAST using different penalty hyper-parameters $\alpha$. Panels correspond to variate.}
\label{supp_fig:Alpha_ISE}
\end{figure}

Figure~\ref{supp_fig:Alpha_ISE} indicates that estimation accuracy of the true underlying smooth functions is consistent across the range of $\alpha$ values tested.

\begin{figure}[h]
\centering
\includegraphics[width=12cm]{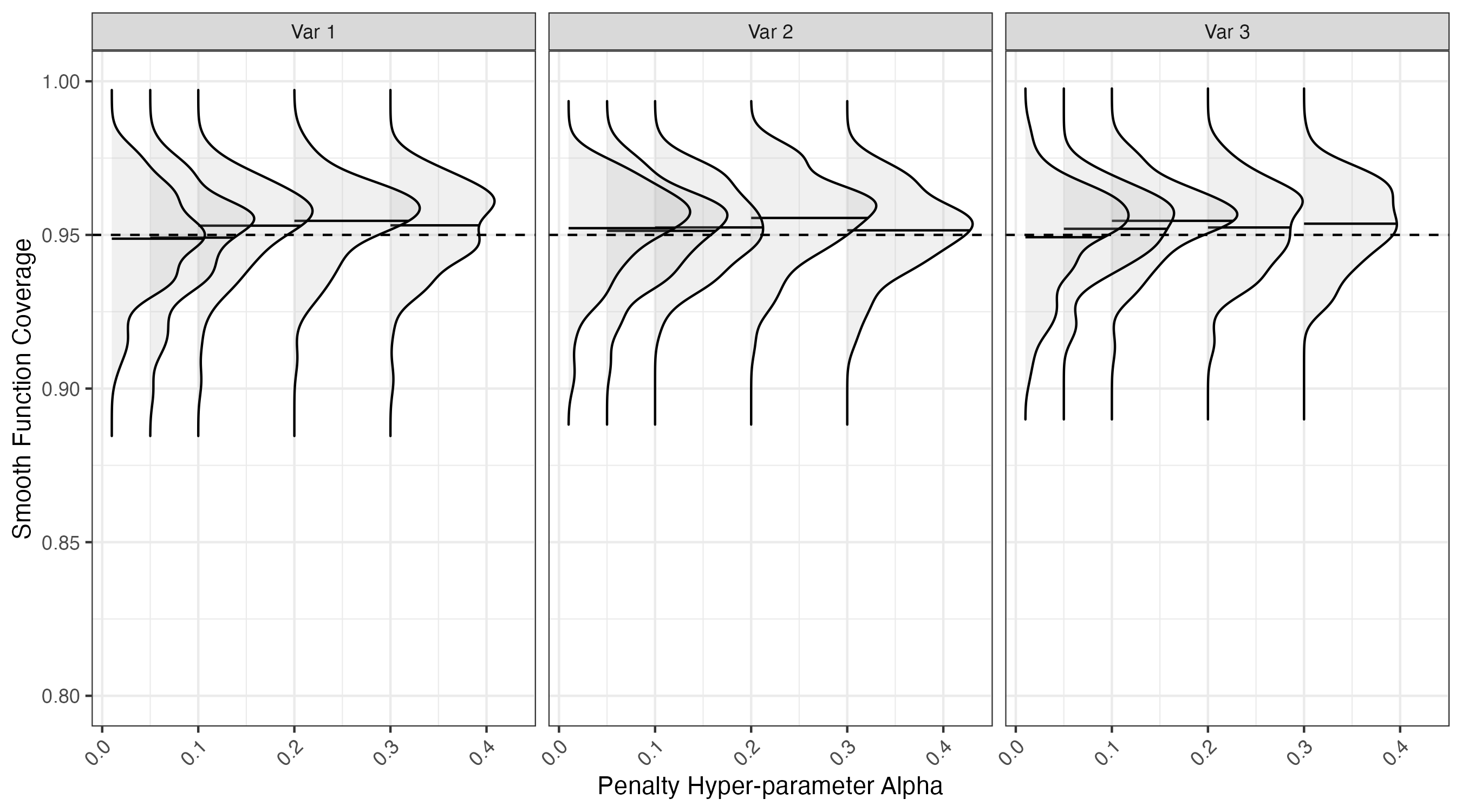}
\caption{Kernel smooths of coverage for the $95\%$ credible intervals of latent smooth functions produced by MSFAST under different penalty hyper-parameters $\alpha$. Panels correspond to variate.}
\label{supp_fig:Alpha_COV}
\end{figure}

Figure~\ref{supp_fig:Alpha_COV} indicates that inference upon the true underlying smooth functions is also consistent across the range of $\alpha$ values tested.

\subsection*{Sensitivity Analysis of FPC Basis Dimension $K$}\label{supp:k_sens}

To evaluate the dependence of MSFAST on the chosen FPC basis dimension $K$, we leveraged the simulation structure in Section~\ref{sec:MSFAST_sim}, fixing $P = 3$, $Q = 20$, $I = 100$, $M = 100$, penalty hyper-parameter $\alpha = 0.1$, and signal-to-noise ratio of $4$, all while varying $K$ between $2$ and $6$. As the true number of eigenfunctions was fixed at $K = 3$, this structure provides some idea of the effects of both under- and over-parameterizing the FPC basis modeled by MSFAST. We evaluated the ability of MSFAST to capture the latent smooth functions under each $K$ value using the estimation and inference metrics introduced in Section~\ref{sec:MSFAST_sim}.

\begin{figure}[h]
\centering
\includegraphics[width=12cm]{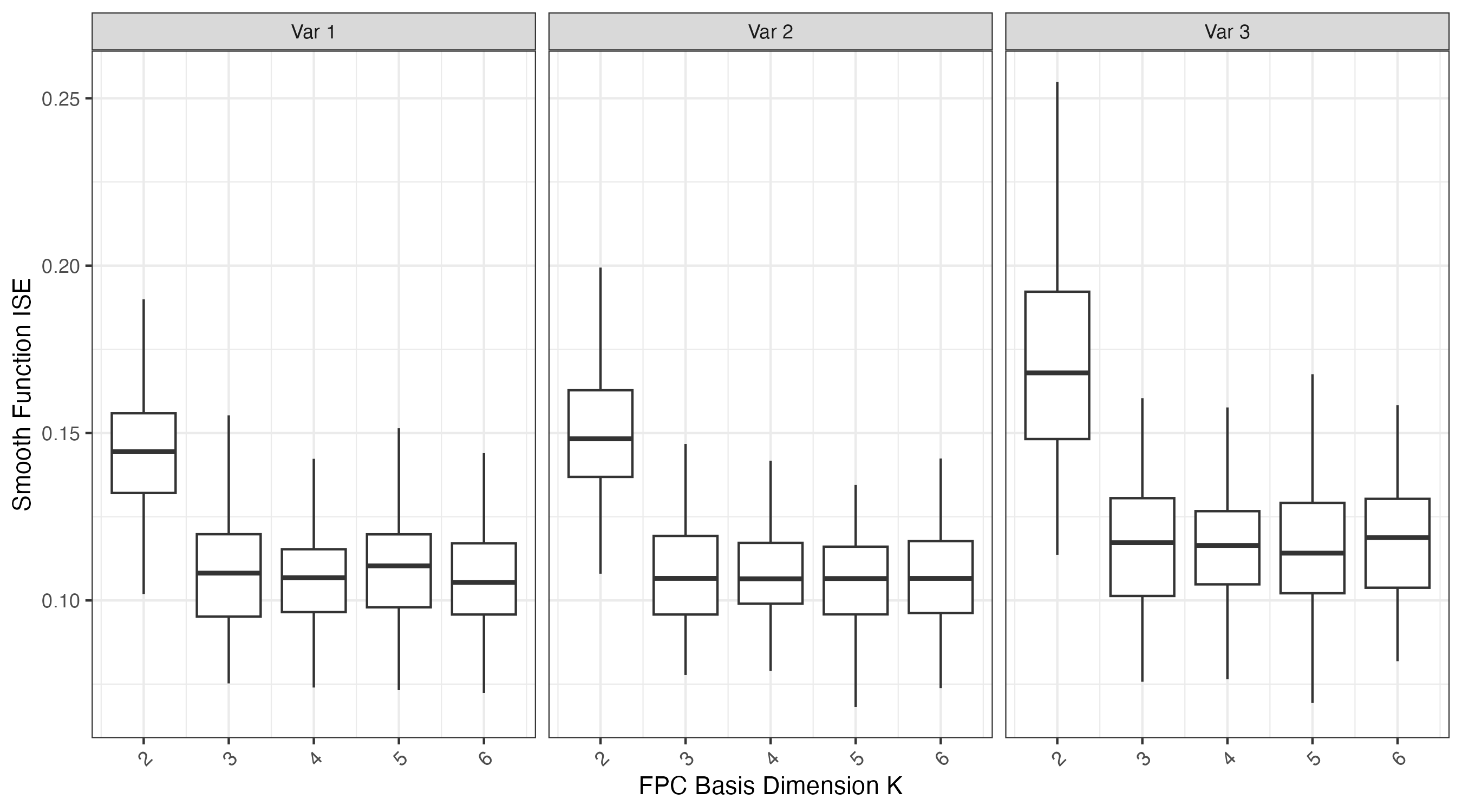}
\caption{Boxplots of RISE from applying MSFAST with different FPC basis dimensions $K$. Panels correspond to variate.}
\label{supp_fig:K_ISE}
\end{figure}

Figure~\ref{supp_fig:K_ISE} indicates that under-specifying the model ($K = 2 < 3$) does increase RISE, though MSFAST is still superior to taking the simple mean. However, over-parameterization has no effect on estimation accuracy, as the model produces identical RISE distributions for all values of $K \geq 3$.

\begin{figure}[h]
\centering
\includegraphics[width=12cm]{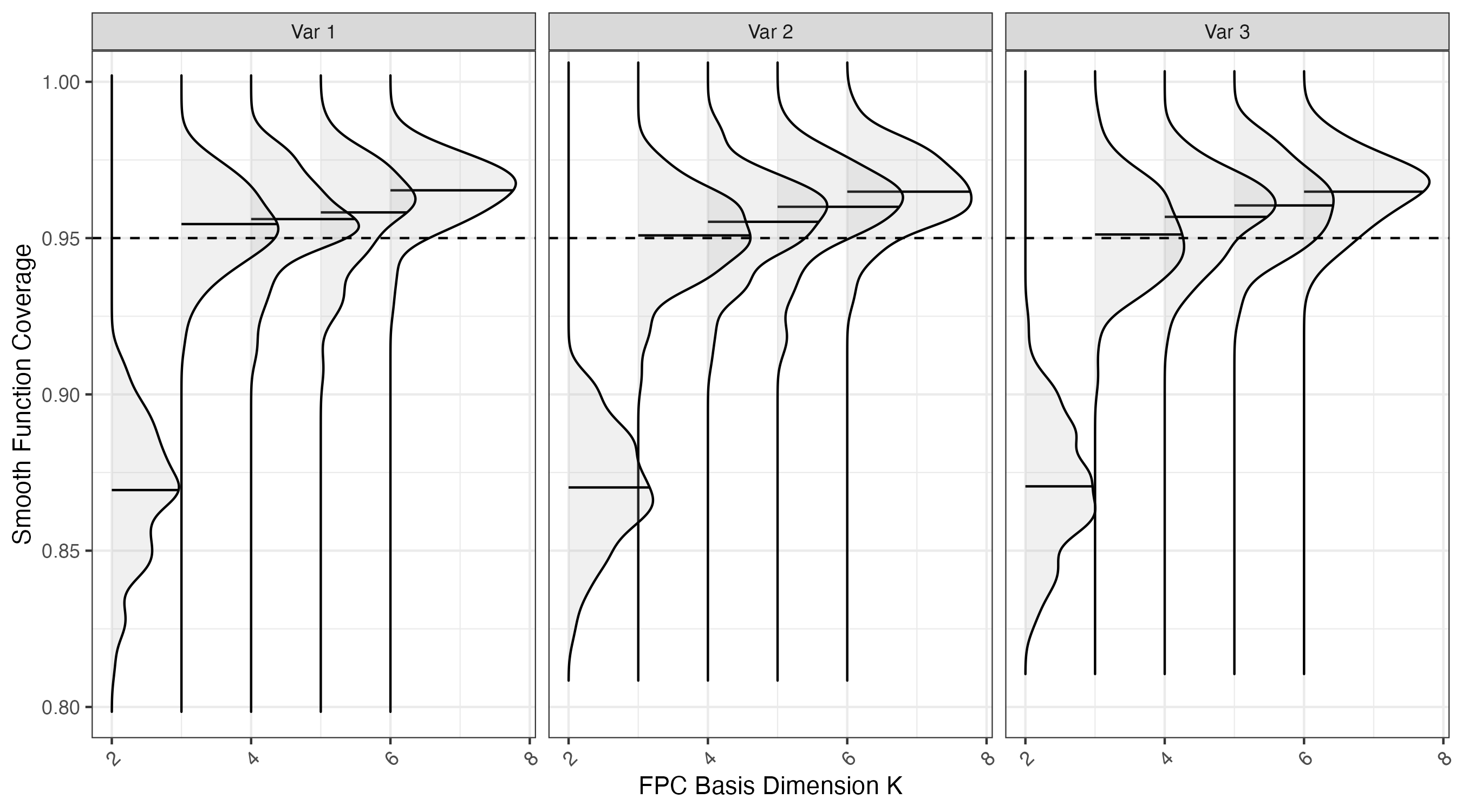}
\caption{Kernel smooths of coverage for the $95\%$ credible intervals of latent smooth functions produced by MSFAST under different FPC basis dimensions $K$. Panels correspond to variate.}
\label{supp_fig:K_COV}
\end{figure}

Inference results from Figure~\ref{supp_fig:K_COV} largely align with the estimation results from Figure~\ref{supp_fig:K_ISE}. Inference is quite similar when the number of FPCs is at least as large as employed in the generative model ($K \geq 3$), but coverage is sub-nominal (close to $\approx 0.85$) when an insufficient number of FPCs are fit, $K = 2$.

\subsection*{Parallelization}\label{supp:parallel}

All STAN models are inherently able to be parallelized by including additional MCMC chains. For this to be employed most effectively, the number of warm-up iterations should be relatively modest, as these samples represent pure overhead. We have found this to be the case for MSFAST, particularly with the use of the posterior alignment routine described in Section~\ref{sec:ACE}. Using this routine, we often find that as little as $\approx 250$ samples can be used as warm-up. So, for a desired number of posterior samples, one can distribute this number over as many chains as they have computational cores, including a small warm-up for each chain. In this way, MSFAST is well-suited to parallel computation through multiple MCMC chains. This approach is also quite simple to implement in modern Bayesian software, as {\ttfamily rstan} and {\ttfamily cmdstanr} both have arguments in their sampling calls to directly specify the number of parallel sampling chains to initialize.

Along with using multiple MCMC chains, we can alternatively parallelize computation by covariate within each chain through multi-threading. This technique is more involved, but we will directly illustrate the necessary changes in {\ttfamily STAN}. The multi-threaded implementation of MSFAST leverages the fact that calculating the likelihood contributions (and corresponding gradients) is embarrassingly parallelizable across covariates. While more resource intensive, this parallelization structure theoretically allows MSFAST to scale nearly identically regardless of the number of modeled covariates, so long as sufficient computational resources are available. This is due to the majority of the computational burden for MSFAST being associated with performing these likelihood/gradient calculations, as indicated by our performance profiling.

The first change from the original {\ttfamily STAN} implementation is the addition of a function within the previously unused {\ttfamily functions} block. This function leverages the {\ttfamily reduce\_sum()} function within {\ttfamily STAN} which enables parallelization over multiple threads. The function, named {\ttfamily partial\_sum\_lik()} can be found below. Notice that it takes as arguments all elements required to calculate those elements of the likelihood which correspond to a particular covariate.

\singlespacing
\begin{lstlisting}[language=Stan]
functions {
  real partial_sum_lik(array[] int p_slice, int start, int end,
                      array[] int Tp_card, array[] int Subj, 
                      array[] int S, matrix Psi, matrix Scores,
                      matrix B, vector w_mu, vector sigma2, 
                      array[] int start_indices,
                      vector Y, int Q, int M, int K) {
    real acc = 0;    // accumulated log density

    for (p in p_slice) {
      int sdx = (p-1)*Q+1;  // Slicing indices by-variable
      int edx = p*Q;
      int Tp = Tp_card[p];  // Size of variable data slice
      
      // Variable subject and time indices (slice)
      array[Tp] int Subj_p = segment(Subj, start_indices[p], Tp);
      array[Tp] int S_p = segment(S, start_indices[p], Tp);

      // FPC matrix, Random Effects, then Fixed Effects
      matrix[Tp, K] Phi_mat = B[S_p, ] * Psi[sdx:edx, ];
      vector[Tp] Theta = rows_dot_product(Scores[Subj_p,], Phi_mat);
      vector[Tp] mu = B[S_p,] * w_mu[sdx:edx];

      // Add likelihood to accumulator
      acc += normal_lpdf(segment(Y, start_indices[p], Tp) | 
                          mu+Theta, sqrt(sigma2[p]));
    }
    return acc;
  }
}
\end{lstlisting}
\doublespacing

Prior to using this function within the {\ttfamily Model} block, we must create some of the required constants using the input data. In particular, the {\ttfamily start\_indices} array (which stores where each covariate begins within the stacked data) must be generated. We also require a simple vector/array from $1$ to $P$ to parallelize over (denoted {\ttfamily p\_vals} here). These are generated within the {\ttfamily transformed data} block as follows.

\singlespacing
\begin{lstlisting}[language=Stan]
transformed data{
  array[P] int start_indices;
  array[P] int p_vals;
  {
    int pos = 1;
    for(p in 1:P){
      p_vals[p] = p;
      start_indices[p] = pos;
      pos = pos + Tp_card[p];
    } 
  }
}
\end{lstlisting}
\doublespacing

The final change which must be made is the implementation of the defined {\ttfamily partial\_sum\_lik()} function through {\ttfamily reduce\_sum()} within the {\ttfamily model} block. We present this section of the {\ttfamily STAN} code, suitably updated.

\singlespacing
\begin{lstlisting}[language=Stan]
model {
  // Variance component priors
  lambda (*$\sim$*) inv_gamma(0.001, 0.001); 
  sigma2 (*$\sim$*) inv_gamma(0.001, 0.001);
  
  // Smoothing priors
  h_mu (*$\sim$*) gamma(0.01, 0.01); 
  H (*$\sim$*) gamma(0.01, 0.01); 
  
  int sx;
  int ex;
  for(p in 1:P){
    sx = (p-1)*Q+1;
    ex = p*Q;
    
    target += Q/2*log(h_mu[p])-h_mu[p]/2*quad_form(P_alpha, w_mu[sx:ex]);
    
    for(k in 1:K){
      target += Q/2*log(H[k])-H[k]/2*quad_form(P_alpha, Psi[sx:ex,k]);
    }
  }
  
  // Normal prior inducing uniform prior on Stiefel Manifold
  to_vector(X) (*$\sim$*) std_normal();
  
  // Priors on unscaled scores 
  to_vector(Xi_Raw) (*$\sim$*) std_normal();
  
  // Model likelihood
  target += reduce_sum(
    partial_sum_lik,      // likelihood over variable slice
    p_vals,               // split by variables
    1,                    // grainsize of 1 (1 variable each)
    Tp_card, Subj, S, Psi, Scores, B, w_mu, sigma2, 
    start_indices, Y, Q, M, K
  );
}
\end{lstlisting}
\doublespacing

This concludes the requisite updates to the original {\ttfamily STAN} code in order to make use of parallel computations over modeled covariates within each sampling chain. Note that, depending upon the {\ttfamily STAN} interface used, some settings may need to be adjusted to enable models to use this type of multi-threading.

\begin{figure}[h]
\centering
\includegraphics[width=14cm]{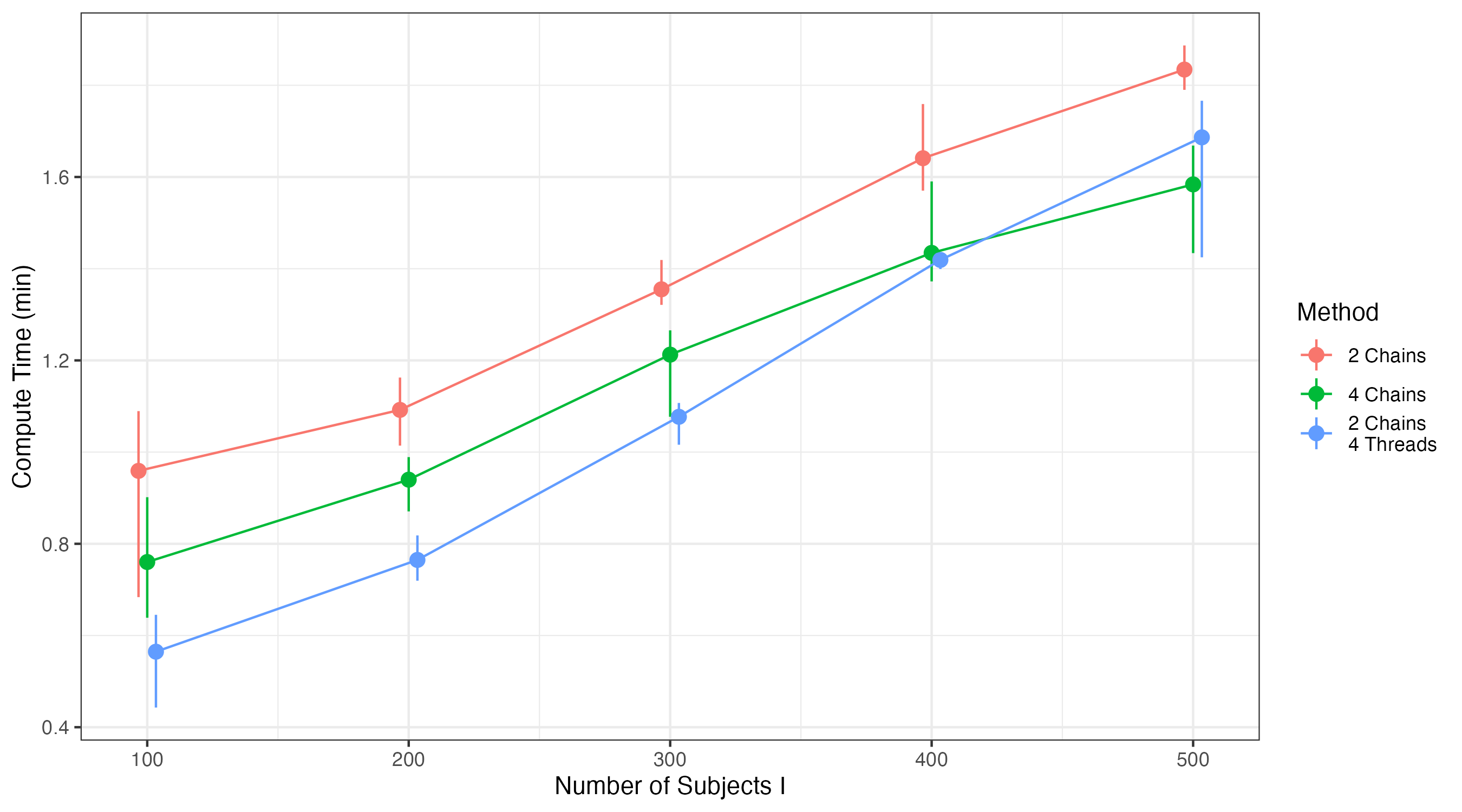}
\caption{Compute time in minutes for MSFAST compared to MSFAST parallelized using both more MCMC chains (4 Chains) and multi-threading within chain (2 Chains 4 Threads).}
\label{supp_fig:Parallel_Comp}
\end{figure}

To evaluate whether each of these parallelization approaches provide sufficient computational benefit to outweigh the associated overhead, we performed an additional timing simulation. We used the exact simulation specification for the timing profiling done in Section~\ref{sec:MSFAST_sim}, though we excluded $I = 1000$, and applied standard MSFAST, MSFAST with the same number of total samples but double the number of MCMC chains, and multi-threaded MSFAST (2 threads per chain). The results, shown in Figure~\ref{supp_fig:Parallel_Comp}, are promising for this simple simulation, with both parallel approaches reducing computation time by between $10\%$ and $40\%$.  This is also a relatively small computational load, as indicated by the compute times. We hypothesize that the performance improvement may be greater for more intensive datasets.

\subsection*{Choice of Basis}\label{supp:bases}

To illustrate the differences between our chosen basis of orthogonalized B-splines \citep{redd_comment_2012}, and the Splinet basis \citep{liu_splinets_2020} chosen to handle densely observed data as part of FAST (see \cite{sartini_fast_2026}), we plot the two bases together for a reasonable dimension $Q = 5$ in Supplemental Figure~\ref{supp_fig:Basis_Comp}.

\begin{figure}[h]
\centering
\includegraphics[width=14cm]{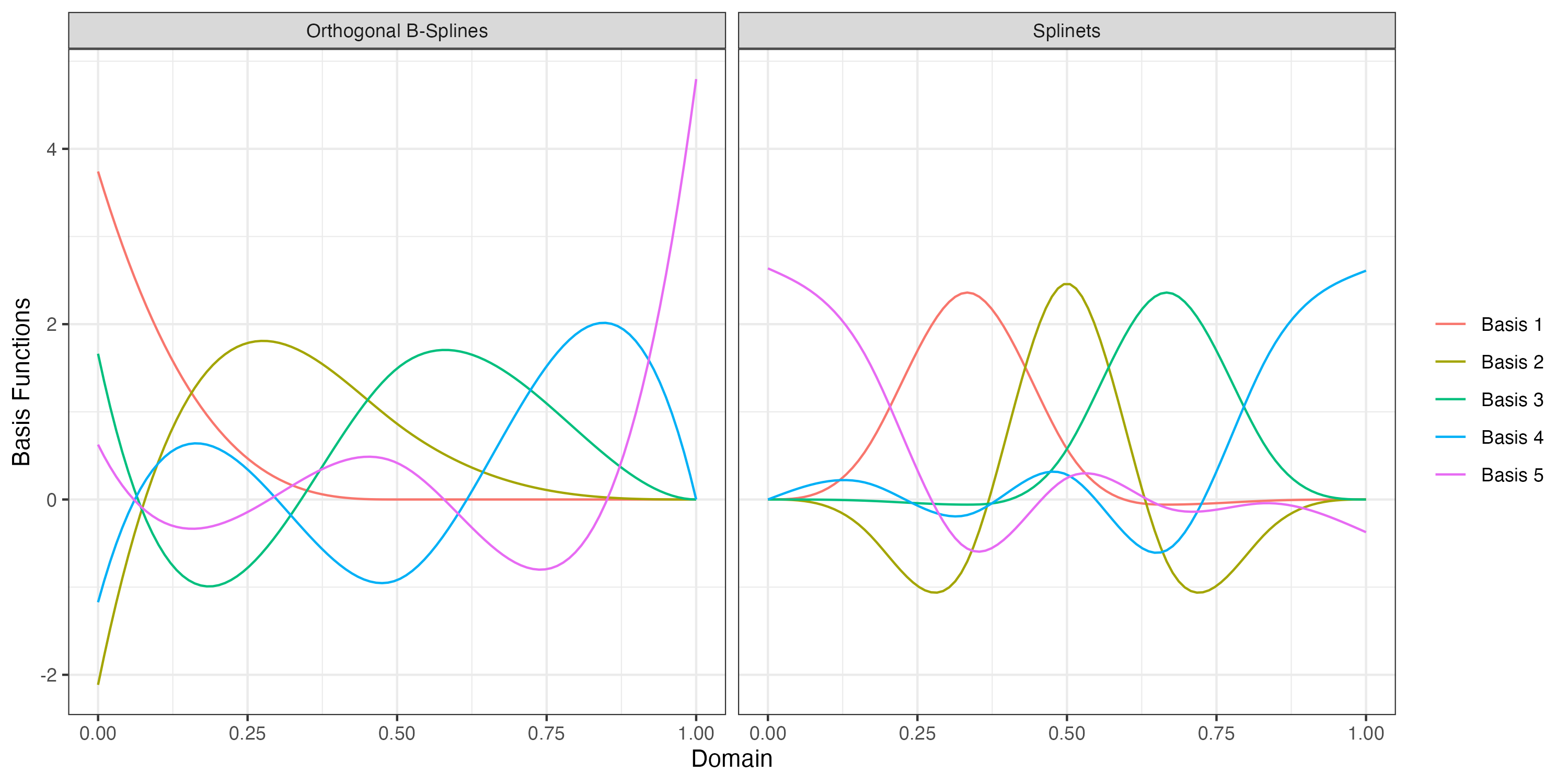}
\caption{The left panel visualizes the elements of a $Q=5$ dimensional orthogonalized B-spline basis, as presented by \cite{redd_comment_2012}. The right panel provides the corresponding $Q=5$ dimensional Splinet basis as proposed by \cite{liu_splinets_2020}, suitably augmented with orthogonalized slope and intercept. In both panels, color indicates basis element.}
\label{supp_fig:Basis_Comp}
\end{figure}

From Supplemental Figure~\ref{supp_fig:Basis_Comp}, we see that nearly all of the Orthogonalized B-splines have non-negligible magnitude over the entire domain. This is in stark contrast to Splinets, which have rather small magnitudes outside of local neighborhoods corresponding to their respective knots. This is by design, as locality of this type can be computationally beneficial in many applications. However, for sparsely-observed data, particularly when certain regions of the domain have very few observations over all individuals, this type of locality can give rise to lack of identifiability in the spline coefficient posterior space. It is for this reason that we choose to use the smoother, less local Orthogonalized B-splines.

\subsection*{Joint and Conditional Posteriors}\label{supp:posteriors}

We begin by providing the likelihood and prior distribution for the full sparse, multivariate data case described in Section~\ref{sec:Multivariate}. This includes the univariate model described in Section~\ref{sec:Univariate} as a special case. For the purposes of notational brevity, we use $\Gamma(a, b)$ to denote the gamma distribution with shape $a$ and rate $b$, $\Gamma^{-1}(a, b)$ to denote the inverse gamma with shape $a$ and scale $b$, $N(\mu, \sigma^2)$ to refer to the normal distribution with mean $\mu$ and variance $\sigma^2$, and $MVN(\boldsymbol{\mu, \Sigma})$ to refer to the multivariate normal distribution with mean vector $\boldsymbol{\mu}$ and variance-covariance $\Sigma$. Throughout, $\mathbf{I}_A$ denotes the identity matrix of dimension $A$, and $[X|Y]$ denotes the distribution of $X$ given $Y$.

\subsubsection*{Variance Component Priors}

Prior distributions for all variance components and smoothing parameters (which can be thought of as inverse variance components) are as follows. Here, $h_{(\mu, p)}$ refers to the smoothing parameter for the mean function of covariate $p$ and $h_k$ refers to the shared smoothing parameter for the FPCs $\phi^{(p)}_k(t)$.
\begin{align*}
    \sigma_{p}^2 & \sim \Gamma^{-1}(\alpha_\sigma, \beta_\sigma) \quad \forall p = 1, \ldots, P\\
    \lambda_k & \sim \Gamma^{-1}(\alpha_\lambda, \beta_\lambda) \quad \forall k = 1, \ldots, K\\
    h_{(\mu,p)} & \sim \Gamma(\alpha_\mu, \beta_\mu) \quad \forall p= 1, \ldots, P\\
    h_{k} & \sim \Gamma(\alpha_\psi, \beta_\psi) \quad \forall k = 1, \ldots, K
\end{align*}
For the variance components, we set relatively uninformative priors using $\alpha_\sigma = \alpha_\lambda = 0.001$ and $\beta_\sigma = \beta_\lambda = 0.001$. For the smoothing parameters, we set $\alpha_\mu = \alpha_\psi = 0.01$ and $\beta_\mu = \beta_\psi = 0.01$ for weakly-informative priors that hedge against computational challenges due to excessive sparsity. As indicated in Result~\ref{rmk:proper_prior}, this choice will produce a proper joint prior over the $\boldsymbol{\psi}_k$ and $h_k$.

\subsubsection*{Fixed Effect Spline Weight Priors}

MSFAST places smoothing priors on the spline weights $\mu^{(p)}(t)$ through a quadratic posterior penalty on the corresponding spline coefficients $w_\mu^{(p)}$. The equivalent prior density on these coefficients, denoted $f(w_{\mu}^{(p)}|h_{(\mu, p)})$, is characterized as follows:
$$f(w_\mu^{(p)}) \propto h_{(\mu,p)}^{\mathbf{R}(\mathbf{P}_\alpha)/2}\exp\left\{-\frac{h_{(\mu, p)}}{2} \left(w_{\mu}^{(p)}\right)^\top \mathbf{P}_\alpha\left(w_{\mu}^{(p)}\right)\right\}$$
where $\mathbf{R}(\cdot)$ refers to the matrix rank and $\mathbf{P}_\alpha$ is the penalty matrix described in Supplement Section~\ref{supp:penalties}.

\subsubsection*{FPC Spline Weight Priors}

MSFAST uses the same smoothing procedure to define the prior belief about each function $\phi_k^{(p)}(t)$, placing a quadratic penalty on the corresponding spline coefficients $\psi_k^{(p)}$. This implies the following definition of the conditional FPC prior $f(\psi_k^{(p)}|h_k)$, up to normalizing constant:
$$f(\psi_k^{(p)}|h_k) \propto h_k^{\mathbf{R}(\mathbf{P}_\alpha)/2} \exp\left\{-\frac{h_k}{2} \left(\psi_k^{(p)}\right)^\top \mathbf{P}_\alpha \left(\psi_k^{(p)}\right)\right\}$$
where again $\mathbf{R}(\cdot)$ is the rank of the argument matrix and $\mathbf{P}_\alpha$ is the penalty matrix corresponding to the chosen orthogonal spline basis. 

Recall that we additionally require that the full matrix of FPC spline weights, $\boldsymbol{\Psi} = [\boldsymbol{\psi}_1|\cdots|\boldsymbol{\psi}_K]$ for $\boldsymbol{\psi}_k = \{\psi_k^{(1)}, \ldots, \psi_k^{(P)}\}^\top$, is orthonormal. This is required for the corresponding multivariate FPCs to be orthonormal with respect to the sum inner product defined in Section~\ref{sec:Multivariate}. Precisely, this means that $\boldsymbol{\Psi} \in \mathcal{V}_{K,PQ}$, where $\mathcal{V}_{K,PQ}$ is the Stiefel Manifold of orthonormal matrices with dimension $PQ \times K$. We correspondingly add to the FPC prior the indicator $\mathbbm{1}(\boldsymbol{\Psi} \in \mathcal{V}_{K,PQ})$ that the FPC spline weight matrix belongs to this manifold.

\subsubsection*{Score Priors}

In keeping with the definition of multivariate FPCA, as defined by \cite{happ_multivariate_2018} through extension of the traditional Kosambi-Karhunen-Loève decomposition \citep{kosambi_statistics_1943, karhunen_uber_1947, loeve_probability_1978}, the score prior distributions are prescribed. It is key to note here that this distribution encompasses cases of both sparsely and densely observed data, as it describes the latent functions. The scores will have distributions $\xi_{ik} \sim N(0, \lambda_k)$ for $i = 1,\ldots, I$ and $k = 1,\ldots, K$.

\subsubsection*{Model Likelihood}

Mirroring the notation of Section~\ref{sec:Multivariate}, the model likelihood of the data given the parameters, denoted $[\mathbf{Y}|{\rm Paramters}]$, has the following form:
\begin{equation*}
    \begin{split}
    & [\mathbf{Y}|{\rm Paramters}] \\
    & = \prod_{p = 1}^P \prod_{i = 1}^I {\rm MVN}\left(Y_i^{(p)}(\mathbf{T}_i^{(p)})|\mu^{(p)}(\mathbf{T}_i^{(p)}) + \sum_{k = 1}^K\xi_{ik} \phi_k^{(p)}(\mathbf{T}_i^{(p)}), \sigma^2_p \mathbf{I}_{J_i^{(p)}}\right)
    \end{split}
\end{equation*}
This likelihood is in the functional form, which is not ideal for deriving the conditional posterior distributions of the latent spline weights. Towards expressing this density in terms of those weights, we introduce some shorthand notation. Let matrix $\mathbf{B}_i^{(p)} \in \mathbb{R}^{J_i^{(p)} \times K}$ represent $\mathbf{B}(\mathbf{T}_i^{(p)})$, the orthonormal spline basis evaluated at the observed points for participant $i$ and covariate $p$. Using this notation, we have the following likelihood form:
\begin{equation*}
    \begin{split}
    & [\mathbf{Y}|{\rm Paramters}] \\
    & = \prod_{p = 1}^P \prod_{i = 1}^I {\rm MVN}\left(Y_i^{(p)}(\mathbf{T}_i^{(p)})|\mathbf{B}_i^{(p)}\left\{w_\mu^{(p)} + \sum_{k = 1}^K\xi_{ik} \psi_k^{(p)}\right\}, \sigma^2_p \mathbf{I}_{J_i^{(p)}}\right)
    \end{split}
\end{equation*}

\subsubsection*{Joint Posterior}

Using the priors and likelihood previously described, the joint posterior density of the modeled parameters given the observed data is defined up to constant of proportionality to be the following. Throughout, we use $f(X|\Theta)$ to refer to the density $f(\cdot)$ with parameters $\Theta$ evaluated at value $X$.
\begin{equation*}
 \begin{split}
    & \prod_{p = 1}^P\Gamma^{-1}(\sigma_p^2|\alpha_\sigma, \beta_\sigma) \times \Gamma(h_{(\mu, p)}|\alpha_\mu, \beta_\mu) \times h_{(\mu,p)}^{\mathbf{R}(\mathbf{P}_\alpha)/2}\exp\left\{-\frac{h_{(\mu, p)}}{2} \left(w_{\mu}^{(p)}\right)^\top \mathbf{P}_\alpha\left(w_{\mu}^{(p)}\right)\right\} \\
    & \times \prod_{i = 1}^I {\rm MVN}\left(Y_i^{(p)}(\mathbf{T}_i^{(p)})|\mathbf{B}_i^{(p)}\left\{w_\mu^{(p)} + \sum_{k = 1}^K\xi_{ik} \psi_k^{(p)}\right\}, \sigma^2_p \mathbf{I}_{J_i^{(p)}}\right) \\
    & \times \prod_{k = 1}^K  N(\xi_{ik}|0, \lambda_k) \Gamma^{-1}(\lambda_k|\alpha_\lambda, \beta_\lambda)  \Gamma(h_k|\alpha_\psi, \beta_\psi) \\
    & \times  h_k^{\mathbf{R}(\mathbf{P}_\alpha)/2} \exp\left\{-\frac{h_k}{2} \left(\psi_k^{(p)}\right)^\top \mathbf{P}_\alpha \left(\psi_k^{(p)}\right)\right\}\times \mathbbm{1}(\boldsymbol{\Psi} \in \mathcal{V}_{K, PQ})
\end{split}
\end{equation*}
We can now use this joint posterior to derive the individual conditional posterior distributions of each modeled parameter.

\subsubsection*{Smoothing Parameter Conditional Posteriors}

We begin with the conditional posterior distribution of the smoothing parameters for the fixed effects over $p = 1,\ldots, P$. 
\begin{align*}
    [h_{(\mu, p)}|{\rm others}] & \propto \Gamma(h_{(\mu, p)}|\alpha_\mu, \beta_\mu) \times h_{(\mu,p)}^{\mathbf{R}(\mathbf{P}_\alpha)/2}\exp\left\{-\frac{h_{(\mu, p)}}{2} \left(w_{\mu}^{(p)}\right)^\top \mathbf{P}_\alpha\left(w_{\mu}^{(p)}\right)\right\}\\
    & \propto h_{(\mu, p)}^{[\alpha_\mu + \mathbf{R}(\mathbf{P}_\alpha)/2] - 1} \exp\left\{-\left(\beta_\mu + \frac{\left(w_{\mu}^{(p)}\right)^\top \mathbf{P}_\alpha\left(w_{\mu}^{(p)}\right)}{2}\right) h_{(\mu, p)}\right\}
\end{align*}
The above has the form of the Gamma distribution. To be precise:
$$[h_{(\mu, p)}|{\rm others}] \sim \Gamma\left(\alpha_\mu + \mathbf{R}[\mathbf{P}_\alpha]/2, \beta_\mu + \frac{\left(w_{\mu}^{(p)}\right)^\top \mathbf{P}_\alpha\left(w_{\mu}^{(p)}\right)}{2}\right)$$

This procedure is very similar for deriving the conditional posterior of the smoothing parameters for the FPCs for each $k = 1, \ldots, K$.
\begin{align*}
    [h_k|{\rm others}] & \propto \Gamma(h_k|\alpha_\psi, \beta_\psi) \times  \prod_{p = 1}^P h_k^{\mathbf{R}(\mathbf{P}_\alpha)/2} \exp\left\{-\frac{h_k}{2} \left(\psi_k^{(p)}\right)^\top \mathbf{P}_\alpha \left(\psi_k^{(p)}\right)\right\}\\
    & \propto h_k^{[\alpha_\psi + P\mathbf{R}(\mathbf{P}_\alpha)/2] - 1} \exp\left\{-\left(\beta_\psi + \sum_{p = 1}^P\frac{\left(\psi_k^{(p)}\right)^\top \mathbf{P}_\alpha \left(\psi_k^{(p)}\right)}{2}\right)h_k\right\}
\end{align*}
This also has the Gamma distributional form; to be precise:
$$[h_k|{\rm others}] \sim \Gamma\left(\alpha_\psi + P\mathbf{R}[\mathbf{P}_{\alpha}]/2, \beta_\psi + \sum_{p = 1}^P\frac{\left(\psi_k^{(p)}\right)^\top \mathbf{P}_\alpha \left(\psi_k^{(p)}\right)}{2}\right)$$

\subsubsection*{Variance Component Conditional Posteriors}

We begin with the conditional posteriors of the noise variances.
\begin{align*}
    [\sigma^2_p|{\rm others}] & \propto \prod_{i = 1}^I {\rm MVN}\left(Y_i^{(p)}(\mathbf{T}_i^{(p)})|\mathbf{B}_i^{(p)}\left\{w_\mu^{(p)} + \sum_{k = 1}^K\xi_{ik} \psi_k^{(p)}\right\}, \sigma^2_p \mathbf{I}_{J_i^{(p)}}\right)\\
    & \quad \times \Gamma^{-1}(\sigma_p^2|\alpha_\sigma, \beta_\sigma)\\
    & \propto \prod_{i = 1}^I (\sigma_p^2)^{-J_i^{(p)}/2} \exp(-\frac{1}{2\sigma^2_p}||Y_i^{(p)}(\mathbf{T}_i^{(p)}) - \mathbf{B}_i^{(p)}\left\{w_\mu^{(p)} + \sum_{k = 1}^K\xi_{ik} \psi_k^{(p)}\right\}||^2)\\
    & \quad \times (\sigma_p^2)^{-\alpha_\sigma - 1}\exp(-\frac{\beta_\sigma}{\sigma_p^2})
\end{align*}
For notational simplicity, we next introduce the $J_i^{(p)}-$dimensional residual vector $\mathbf{R}_i^{(p)} = Y_i^{(p)}(\mathbf{T}_i^{(p)}) - \mathbf{B}_i^{(p)}\left\{w_\mu^{(p)} + \sum_{k = 1}^K\xi_{ik} \psi_k^{(p)}\right\}$. Using this quantity, the derivation can proceed as follows.
\begin{align*}
    [\sigma^2_p|{\rm others}] & \propto \prod_{i = 1}^I (\sigma_p^2)^{-J_i^{(p)}/2} \exp(-\frac{1}{2\sigma^2_p}||\mathbf{R}_i^{(p)}||^2) \times (\sigma_p^2)^{-\alpha_\sigma - 1}\exp(-\frac{\beta_\sigma}{\sigma_p^2})\\
    & \propto (\sigma_p^2)^{-\sum_{i = 1}^I J_i^{(p)}/2} \exp(-\sum_{i = 1}^I\frac{||\mathbf{R}_i^{(p)}||^2}{2\sigma^2_p}) \times (\sigma_p^2)^{-\alpha_\sigma - 1}\exp(-\frac{\beta_\sigma}{\sigma_p^2}) \\
    & \propto (\sigma_p^2)^{-(\alpha_\sigma + \sum_{i = 1}^I J_i^{(p)}/2) - 1} \exp(-\left\{\beta_\sigma + \frac{\sum_{i = 1}^I ||\mathbf{R}_i^{(p)}||^2}{2}\right\}/\sigma^2_p)
\end{align*}
The resulting functional form is equivalent to the inverse Gamma distribution.
$$[\sigma_p^2|{\rm others}] \sim \Gamma^{-1}\left(\alpha_\sigma + \sum_{i = 1}^I J_i^{(p)}/2,\beta_\sigma  + \frac{\sum_{i = 1}^I ||\mathbf{R}_i^{(p)}||^2}{2} \right)$$

We can next derive the conditional posterior distribution for the eigenvalues $\lambda_1, \ldots, \lambda_K$. 
\begin{align*}
    [\lambda_k|{\rm others}] & \propto  \Gamma^{-1}(\lambda_k|\alpha_\lambda, \beta_\lambda) \prod_{i = 1}^I N(\xi_{ik}|0, \lambda_k) \\
    & \propto  \lambda_k^{-I/2}\exp\left(-\frac{\sum_{i = 1}^I \xi_{ik}^2}{2\lambda_k}\right) \times \lambda_k^{-\alpha_\lambda - 1} \exp\left(-\frac{\beta_\lambda}{ \lambda_k}\right) \\
    & \propto (\lambda_k)^{-(I/2 + \alpha_\lambda) - 1} \exp\left(-\left\{\frac{\sum_{i = 1}^I \xi_{ik}^2}{2} + \beta_\lambda\right\}/\lambda_k\right)
\end{align*}
This joint distribution has the form of independent inverse Gamma distributions for each $\lambda_k$: $[\lambda_k|{\rm others}] \sim \Gamma^{-1}(I/2 + \alpha_\lambda, \frac{\sum_{i = 1}^I \xi_{ik}^2}{2} + \beta_\lambda)$. 

\subsubsection*{Score Conditional Posteriors}\label{supp:Score_Posterior}

We derive the conditional posterior for the set of participant-specific scores $\boldsymbol{\xi}_{i} = \{\xi_{i1}, \ldots, \xi_{iK}\} \in \mathbb{R}^K$, as this distribution is central to the performance of efficient dynamic prediction as described in Section~\ref{sec:prediction}. For this derivation, we will use the same notational shorthands introduced in Result~\ref{Thm:Scores}: letting $\mathbf{R}_i^{(p)} = Y_i^{(p)}(\mathbf{T}_i^{(p)}) - \mathbf{B}(\mathbf{T}_i^{(p)}) \mathbf{w}_\mu^{(p)} = Y_i^{(p)}(\mathbf{T}_i^{(p)}) - \mathbf{B}_i^{(p)} \mathbf{w}_\mu^{(p)} $ refer to the $J_i^{(p)}$-dimensional fixed effects residuals, $\boldsymbol{\Psi}^{(p)} = [\psi_1^{(p)}|\cdots|\psi_K^{(p)}] \in \mathbb{R}^{Q \times K}$, and $\boldsymbol{\Lambda}$ indicate the diagonal matrix of eigenvalues $\lambda_k$.
\begin{align*}
    & [\boldsymbol{\xi}_i|{\rm others}] \\
    & \propto \prod_{p = 1}^P {\rm MVN}\left(Y_i^{(p)}(\mathbf{T}_i^{(p)})|\mathbf{B}_i^{(p)}\left\{w_\mu^{(p)} + \sum_{k = 1}^K\xi_{ik} \psi_k^{(p)}\right\}, \sigma^2_p \mathbf{I}_{J_i^{(p)}}\right) \times {\rm MVN}(\boldsymbol{\xi}_i|\vec{0}, \boldsymbol{\Lambda})\\
    & \propto \exp\left(-\sum_{p = 1}^P\frac{||\mathbf{R}_i^{(p)} - \mathbf{B}_i^{(p)}\boldsymbol{\Psi}^{(p)} \boldsymbol{\xi}_i||^2}{2\sigma_p} \right) \times \exp\left(-\frac{\boldsymbol{\xi}_i^\top \boldsymbol{\Lambda}^{-1} \boldsymbol{\xi}_i}{2}\right)\\
    & \propto \exp\left(-\sum_{p = 1}^P \frac{-2[\mathbf{R}_i^{(p)}]^\top\mathbf{B}_i^{(p)}\boldsymbol{\Psi}^{(p)}\boldsymbol{\xi}_i + ||\mathbf{B}_i^{(p)}\boldsymbol{\Psi}^{(p)}\boldsymbol{\xi}_i||^2}{2\sigma_p^2} - \frac{\boldsymbol{\xi}_i^\top \boldsymbol{\Lambda}^{-1} \boldsymbol{\xi}_i}{2}\right)\\
    & \propto \exp\left(-\frac{1}{2}\left\{-2\sum_{p = 1}^P\frac{[\mathbf{R}_i^{(p)}]^\top \mathbf{B}_i^{(p)}\boldsymbol{\Psi}^{(p)}}{\sigma_p^2}\boldsymbol{\xi}_i + \boldsymbol{\xi}_i^\top\left[\sum_{p = 1}^P\frac{(\boldsymbol{\Psi}^{(p)})^\top(\mathbf{B}_i^{(p)})^\top\mathbf{B}_i^{(p)}\boldsymbol{\Psi}^{(p)}}{\sigma_p^2} + \boldsymbol{\Lambda}^{-1}\right]\boldsymbol{\xi}_i \right\}\right)
\end{align*}
Completing the square above, we find a multivariate normal distribution with variance-covariance $\boldsymbol{\Sigma} = \left[\sum_{p = 1}^P\frac{(\boldsymbol{\Psi}^{(p)})^\top(\mathbf{B}_i^{(p)})^\top\mathbf{B}_i^{(p)}\boldsymbol{\Psi}^{(p)}}{\sigma_p^2} + \boldsymbol{\Lambda}^{-1}\right]^{-1}$ and mean $\mu = \boldsymbol{\Sigma} \left[\sum_{p = 1}^P\frac{[\mathbf{R}_i^{(p)}]^\top \mathbf{B}_i^{(p)}\boldsymbol{\Psi}^{(p)}}{\sigma_p^2}\right]^\top$. Note that we should always be able to calculate $\boldsymbol{\Sigma}$ through inversion due to the diagonal entries of $\boldsymbol{\Lambda}$ being strictly positive.

\subsubsection*{Fixed Effect Spline Weight Posteriors}

We derive the conditional posterior of $w_\mu^{(p)}$ for $p = 1, \ldots, P$, as follows.
\begin{align*}
    & [w_\mu^{(p)}|{\rm others}] \\
    & \propto \exp\left\{-\frac{h_{(\mu, p)}}{2} \left(w_\mu^{(p)}\right)^\top \mathbf{P}_\alpha \left(w_\mu^{(p)}\right)\right\} \\
    & \quad \times \prod_{i = 1}^I {\rm MVN}\left(Y_i^{(p)}(\mathbf{T}_i^{(p)})|\mathbf{B}_i^{(p)}\left\{w_\mu^{(p)} + \sum_{k = 1}^K\xi_{ik} \psi_k^{(p)}\right\}, \sigma^2_p \mathbf{I}_{J_i^{(p)}}\right)\\
    & \propto \exp\left\{-\frac{h_{(\mu, p)}}{2} \left(w_\mu^{(p)}\right)^\top \mathbf{P}_\alpha \left(w_\mu^{(p)}\right)\right\} \\
    & \quad \times \exp\left\{-\frac{1}{2\sigma_p^2}\sum_{i = 1}^I||Y_i^{(p)}(\mathbf{T}_i^{(p)}) - \mathbf{B}_i^{(p)}\left(w_\mu^{(p)} + \sum_{k = 1}^K\xi_{ik} \psi_k^{(p)}\right)||^2\right\}
\end{align*}
For ease of notation, we now introduce the shorthand $\mathbf{D}_i^{(p)} = Y_i^{(p)}(\mathbf{T}_i^{(p)}) - \mathbf{B}_i^{(p)}\sum_{k = 1}^K\xi_{ik} \psi_k^{(p)} \in \mathbb{R}^{J_i^{(p)}}$ for the residual between the observed data and the cumulative impact of the FPCs on covariate $p$. Continuing with this notation:

\begin{align*}
    & [w_\mu^{(p)}|{\rm others}] \\
    & \propto \exp\left\{-\frac{1}{2\sigma_p^2}\sum_{i = 1}^I||\mathbf{D}_i^{(p)} - \mathbf{B}_i^{(p)}w_\mu^{(p)}||^2\right\}\\
    & \quad \times \exp\left\{-\frac{h_{(\mu, p)}}{2} \left(w_\mu^{(p)}\right)^\top \mathbf{P}_\alpha \left(w_\mu^{(p)}\right)\right\} \\
    & \propto \exp\left\{-\frac{1}{2\sigma_p^2}\sum_{i = 1}^I\left(-2[\mathbf{D}_i^{(p)}]^\top\mathbf{B}_i^{(p)}w_\mu^{(p)} + ||\mathbf{B}_i^{(p)}w_\mu^{(p)}||^2 \right)\right\}\\
    & \quad \times \exp\left\{-\frac{h_{(\mu, p)}}{2} \left(w_\mu^{(p)}\right)^\top \mathbf{P}_\alpha \left(w_\mu^{(p)}\right)\right\} \\
    & \propto \exp\left\{-\frac{1}{2}\left(-2\left[\sum_{i = 1}^I\frac{\{\mathbf{D}_i^{(p)}\}^\top\mathbf{B}_i^{(p)}}{\sigma_p^2}\right]w_\mu^{(p)} + [w_\mu^{(p)}]^\top\left[\sum_{i = 1}^I\frac{\{\mathbf{B}_i^{(p)}\}^\top\mathbf{B}_i^{(p)}}{\sigma^2_p}\right]w_\mu^{(p)}\right) \right\}\\
    & \quad \times \exp\left\{-\frac{1}{2} \left(w_\mu^{(p)}\right)^\top \left[h_{(\mu, p)}\mathbf{P}_\alpha \right]\left(w_\mu^{(p)}\right)\right\} \\
    & \propto \exp\Bigg\{-\frac{1}{2}\Bigg(\Big[w_\mu^{(p)}\Big]^\top \Bigg[h_{(\mu, p)}\mathbf{P}_\alpha + \sum_{i = 1}^I\frac{\{\mathbf{B}_i^{(p)}\}^\top\mathbf{B}_i^{(p)}}{\sigma^2_p}\Bigg]\Big[w_\mu^{(p)}\Big] \\
    & \quad - 2\Bigg[\sum_{i = 1}^I\frac{\{\mathbf{D}_i^{(p)}\}^\top\mathbf{B}_i^{(p)}}{\sigma_p^2}\Bigg]w_\mu^{(p)}\Bigg)\Bigg\}
\end{align*}
Completing the square from the above expression yields that $[w_\mu^{(p)}|{\rm others}]$ has multivariate normal distribution with variance-covariance matrix $\boldsymbol{\Sigma} = \left[h_{(\mu, p)}\mathbf{P}_\alpha + \sum_{i = 1}^I\frac{\{\mathbf{B}_i^{(p)}\}^\top\mathbf{B}_i^{(p)}}{\sigma^2_p}\right]^{-1}$ and mean $\boldsymbol{\mu} = \boldsymbol{\Sigma}\left[\sum_{i = 1}^I\frac{\{\mathbf{D}_i^{(p)}\}^\top\mathbf{B}_i^{(p)}}{\sigma_p^2}\right]^\top$.

\subsubsection*{FPC Spline Weight Posteriors}

Finally, we derive the joint posterior distribution for the matrix $\boldsymbol{\Psi} \in \mathbb{R}^{PQ \times K}$ of FPC spline coefficients. We use some notation from Supplement Section~\ref{supp:Score_Posterior}: letting $\boldsymbol{\xi}_i = \{\xi_{i1}, \ldots, \xi_{iK}\}^\top$ refer to the $K-$dimensional score vector and $\mathbf{R}_i^{(p)}$ the $J_i^{(p)}-$dimensional residual $Y_i^{(p)}(\mathbf{T}_i^{(p)}) - \mathbf{B}_i^{(p)}w_\mu^{(p)}$.
\begin{align*}
    [\boldsymbol{\Psi}|{\rm others}] & \propto \prod_{p = 1}^P \prod_{i = 1}^I {\rm MVN}\left(Y_i^{(p)}(\mathbf{T}_i^{(p)})|\mathbf{B}_i^{(p)}\left\{w_\mu^{(p)} + \sum_{k = 1}^K\xi_{ik} \psi_k^{(p)}\right\}, \sigma^2_p \mathbf{I}_{J_i^{(p)}}\right)\\
    & \times \prod_{k = 1}^K \exp\left\{-\frac{h_k}{2} \left(\psi_k^{(p)}\right)^\top \mathbf{P}_\alpha \left(\psi_k^{(p)}\right)\right\} \times \mathbbm{1}(\boldsymbol{\Psi} \in \mathcal{V}_{K, PQ})\\
    & \propto \exp\Bigg\{-\frac{1}{2}\Bigg(\sum_{p = 1}^P \sum_{i = 1}^I \frac{||\mathbf{R}_i^{(p)} - \mathbf{B}_i^{(p)}  \boldsymbol{\Psi}^{(p)}\boldsymbol{\xi}_{i}||^2}{\sigma_p^2} \\
    & \quad + \sum_{p = 1}^P \sum_{k = 1}^K \left[\psi_k^{(p)}\right]^\top h_k\mathbf{P}_\alpha \left[\psi_k^{(p)}\right]\Bigg)\Bigg\} \times \mathbbm{1}(\boldsymbol{\Psi} \in \mathcal{V}_{K, PQ})
\end{align*}
We now make use of the trace matrix operator, ${\rm tr}(A) = \sum_i A_{ii}$, to simplify the above form. We also introduce some additional shorthand matrices. First, diagonal matrix of smoothing parameters $\mathbf{H} = {\rm diag}(\{h_{(1,p)}, \ldots h_k\})$. Next, the matrix of scores $\boldsymbol{\Xi} \in \mathbb{R}^{K \times I}$, where row $i$ corresponds to $\boldsymbol{\xi}_i$. Finally, we introduce the matrices of residual projection onto the chosen orthogonal basis for covariate $p$, $\mathbf{W}^{(p)} \in \mathbb{R}^{I \times K}$ defined such that the $i^{\rm th}$ row of this matrix is equal to $[\mathbf{R}_i^{(p)}]^\top \mathbf{B}_i^{(p)}$. We additionally use ${\rm etr}(\cdot)$ to refer to the exponential trace operator, ${\rm etr}(\mathbf{A}) = \exp({\rm tr}(\mathbf{A}))$ for matrix $\mathbf{A}$. Using these notations:
\begin{align*}
    & [\boldsymbol{\Psi}|{\rm others}]\\
    & \propto \exp\Bigg\{-\frac{1}{2}\Bigg(\frac{1}{\sigma_p^2}\sum_{p = 1}^P \sum_{i = 1}^I\frac{ -2[\mathbf{R}_i^{(p)}]^\top\mathbf{B}_i^{(p)}  \boldsymbol{\Psi}^{(p)}\boldsymbol{\xi}_{i} + ||\mathbf{B}_i^{(p)}  \boldsymbol{\Psi}^{(p)}\boldsymbol{\xi}_{i}||^2}{\sigma_p^2} \\
    & \quad + \sum_{p = 1}^P {\rm tr}\left[ \mathbf{H}\{\boldsymbol{\Psi}^{(p)}\}^\top\mathbf{P}_\alpha\boldsymbol{\Psi}^{(p)}\right ]\Bigg)\Bigg\} \times \mathbbm{1}(\boldsymbol{\Psi} \in \mathcal{V}_{K, PQ})\\
    & \propto \exp\left\{-\frac{1}{2} \left(-\frac{2}{\sigma_p^2}\sum_{p = 1}^P {\rm tr}\left[\mathbf{W}^{(p)} \boldsymbol{\Psi}^{(p)}\boldsymbol{\Xi}\right] + \frac{1}{\sigma_p^2}\sum_{p = 1}^P {\rm tr}\left[\boldsymbol{\Xi}^\top\{\boldsymbol{\Psi}^{(p)}\}^\top \{\mathbf{B}_i^{(p)}\}^\top\mathbf{B}_i^{(p)}\boldsymbol{\Psi}^{(p)} \boldsymbol{\Xi} \right] \right) \right\}\\
    & \quad \times \exp\left\{-\frac{1}{2}\left( \sum_{p = 1}^P {\rm tr}\left[ \mathbf{H}\{\boldsymbol{\Psi}^{(p)}\}^\top\mathbf{P}_\alpha\boldsymbol{\Psi}^{(p)}\right ]\right)\right\} \times \mathbbm{1}(\boldsymbol{\Psi} \in \mathcal{V}_{K, PQ})\\
    & \propto {\rm etr}\Bigg\{\sum_{p = 1}^P \Bigg(\frac{\boldsymbol{\Xi}\mathbf{W}^{(p)}\boldsymbol{\Psi}^{(p)}}{\sigma_p^2} - \frac{\boldsymbol{\Xi}\boldsymbol{\Xi}^\top\{\boldsymbol{\Psi}^{(p)}\}^\top \{\mathbf{B}_i^{(p)}\}^\top\mathbf{B}_i^{(p)}\boldsymbol{\Psi}^{(p)} }{2\sigma_p^2} \\
    & \quad -\frac{\mathbf{H}\{\boldsymbol{\Psi}^{(p)}\}^\top\mathbf{P}_\alpha\boldsymbol{\Psi}^{(p)}}{2}\Bigg)\Bigg\} \times \mathbbm{1}(\boldsymbol{\Psi} \in \mathcal{V}_{K, PQ})
\end{align*}
The above is does not adhere to any known distributional forms on the corresponding Stiefel Manifold $\mathcal{V}_{K, PQ}$. Given this fact, there are no previously published techniques for efficiently sampling from such a distribution, reinforcing the need for a generalized framework such as Hamiltonian Monte Carlo, and an efficient sampling routine using something like parameter expansion, to properly sample $\boldsymbol{\Psi}$.

\subsection*{Dynamic Prediction using Conditional Score Posteriors}\label{supp:cond_pred}

To exemplify how the results in Supplement Section~\ref{supp:posteriors} can be used to perform dynamic predictions conditioned on the population-level covariance and mean estimates, we provide pseudo-code in Algorithm~\ref{alg:score_samp} illustrating the prediction of subject-specific scores $\boldsymbol{\xi}_i$ based upon available data. 

\begin{algorithm}
\caption{Score Sampling Algorithm}\label{alg:score_samp}
\KwData{Number of posterior samples $N$; Number of covariates $P$;\\
\quad $Y_i^{(p)}(\mathbf{T}_i^{(p)}) \in \mathbb{R}^{J_i^{(p)}}$ for $p = 1,\ldots,P$; \quad $\mathbf{B}_i^{(p)} \in \mathbb{R}^{J_i^{(p)} \times K}$ for $p = 1,\ldots,P$;\\
\quad $[\sigma_p^{2}]_n$ posterior noise variance samples for $p = 1, \ldots, P$, $n \leq N$;\\
\quad $[\boldsymbol{\Lambda}]_n = [\diag(\lambda_k)]_n$ posterior eigenvalue samples for $n \leq N$;\\
\quad $[w_\mu^{(p)}]_n$ posterior fixed effect spline samples for $p = 1, \ldots, P$, $n \leq N$;\\
\quad $[\boldsymbol{\Psi}^{(p)}]_n$ posterior FPC spline samples for $p = 1, \ldots, P$, $n \leq N$;\\
}
\KwResult{Scores Samples $[\boldsymbol{\xi}_i]_n$ for $n \leq N$}
\For{$n\gets1$ \KwTo $N$ \KwBy $1$}{
    $\boldsymbol{\Sigma^{-1}} \gets [\boldsymbol{\Lambda}^{-1}]_n$\;
    $\mathbf{M} \gets \boldsymbol{0}$\;
    \For{$p\gets1$ \KwTo $P$ \KwBy $1$}{
        $\boldsymbol{\Sigma}^{-1} \gets \boldsymbol{\Sigma}^{-1} + \frac{[\boldsymbol{\Psi}^{(p)}]_n^\top [\mathbf{B}_i^{(p)}]^\top \mathbf{B}_i^{(p)}[\boldsymbol{\Psi}^{(p)}]_n}{[\sigma_p^2]_n}$\;
        $\mathbf{M} \gets \mathbf{M} + \frac{\{Y_i^{(p)}(\mathbf{T}_i^{(p)}) - \mathbf{B}_i^{(p)}[w_\mu^{(p)}]_n\}^\top\mathbf{B}_i^{(p)}[\boldsymbol{\Psi}^{(p)}]_n}{[\sigma_p^2]_n}$\;
    }
    $\boldsymbol{\mu} = \boldsymbol{\Sigma} \mathbf{M}^\top$\;
    $[\boldsymbol{\xi}_i]_n \sim {\rm MVN}(\boldsymbol{\mu}, \boldsymbol{\Sigma})$\;
}
\end{algorithm}

This technique takes as inputs the observed data $Y_i^{(p)}(\mathbf{T}_i^{(p)})$, basis evaluated at the observed time points $\mathbf{B}_i^{(p)}$, and the posterior samples of $\mathbf{w}_\mu$, FPC weights $\boldsymbol{\Psi}$, eigenvalues $\lambda_k$, and the noise variances $\sigma_p^2$. For each posterior sample, we calculate the multivariate normal posterior mean and variance of the scores $\boldsymbol{\xi}_{i} = \{\xi_{i1}, \ldots, \xi_{iK}\}^\top$ directly using the forms derived in Supplement Section~\ref{supp:posteriors}. We finally sample from this distribution using any sampler of the multivariate normal distribution, returning the sample as an estimate of the scores.

\subsection*{Compute Time Sensitivity in $Q$ and $K$}\label{supp:timing_sens}

To evaluate the computational impact upon $MSFAST$ of changing the spline basis dimension $Q$ and FPC basis dimension $K$, we varied both values over grids ($Q \in \{5, 10, 20, 30, 40\}$ and $K \in \{3, 4, 5, 6\}$) and evaluated the compute time taken for each pair using the timing simulation of Section~\ref{sec:MSFAST_sim}. 

\begin{table}[h]
    \centering
    \begin{tabular}{|c|c|c|c|c|c|}
        \hline
        K/Q & 5     & 10   & 20    & 30    & 40 \\ \hline
        3   & 61.2  & 26.3 & 40.5  & 53.0  & 61.7  \\ \hline
        4   & 103.8 & 46.6 & 61.9  & 80.4 & 93.0 \\ \hline
        5   & 126.5 & 70.7 & 90.6  & 115.3 & 134.1 \\ \hline
        6   & 152.2 & 82.5 & 108.4 & 129.9 & 148.7 \\ \hline
    \end{tabular}
    \caption{Timing sensitivity analysis for MSFAST varying spline basis dimension $Q$ and FPC basis dimension $K$. Reported times are in seconds. $Q$ changes over columns, $K$ over rows.}
    \label{tab:time_table}
\end{table}
Table~\ref{tab:time_table} indicates that choosing too small of a basis ($Q = 5$) has deleterious effects on the computation, resulting in longer computation times. However, for $Q \geq 10$, compute time increases modestly as $Q$ is made larger. Similar patterns of modest increase in compute time are found as $K$ is made larger.
 
\subsection*{Supplemental Simulation Results}\label{supp:sims}

In this section, we present and detail additional results related to the credible interval coverage of functional model components ($\mu^{(p)}(t)$ and $\phi_k^{(p)}(t)$ for $p = 1,\ldots, P$ and $k = 1,\ldots,K$) and to computational efficiency.

We evaluated the validity of inferences drawn upon the functional components $\mu^{(p)}(t), \phi_k^{(p)}(t)$ for $p = 1,\ldots, P$ and $k = 1,\ldots, K$ by assessing the coverage of pointwise 95\% credible intervals over simulation replicates. Coverage was estimated using the same procedure for assessing coverage of the latent smooth functions $Y_{i, {\rm true}}^{(p)}(t)$ (Section~\ref{sec:MSFAST_sim}), now aggregating just over time points $t$ and simulations $b$. As none of the comparator methods were able to perform uncertainty quantification for these functional model components, we treated this evaluation as a pure validation of the inferences drawn by MSFAST. Notably, the mFPCA method has an option to estimate FPC confidence intervals via bootstrap, but we were unable to use this option in testing due to errors which were traced to the PACE package. The resulting distributions, with corresponding means marked using horizontal lines, can be found in the ridge plot of Supplemental Figure~\ref{supp_fig:Multi_Func_Cov}.

\begin{figure}[h]
\centering
\includegraphics[width=12cm]{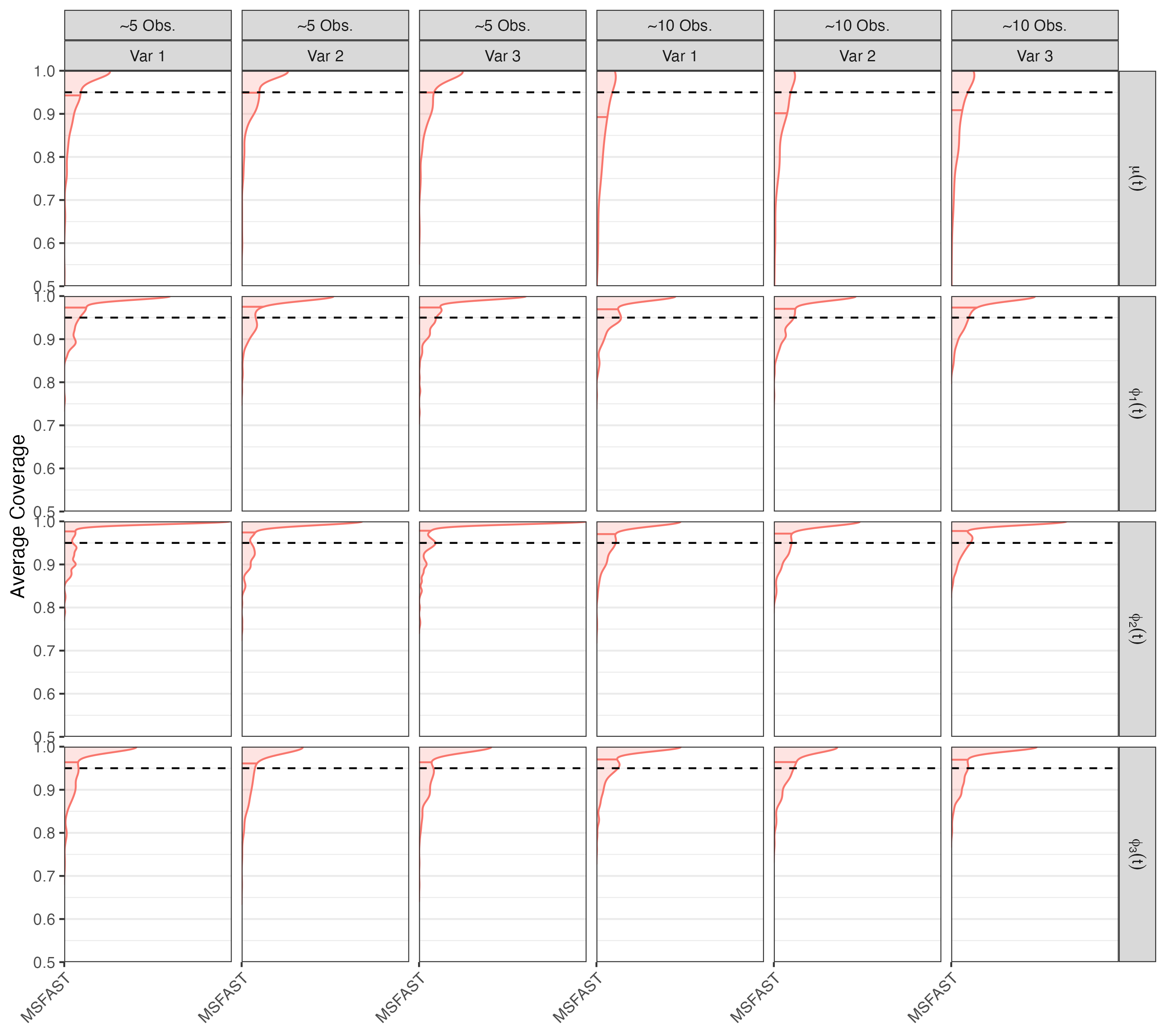}
\caption{Kernel smoother of 95\% credible interval coverage probabilities of the true FPC and mean functions for MSFAST, the only method able to produce intervals for these estimands. Columns $1$-$3$: $5$ expected observations; columns $4$-$6$: $10$ expected observations. Columns $1$, $4$: covariate 1; columns $2$, $5$: covariate 2; columns $3$, $6$: covariate 3. Rows correspond to the means, $\mu^{(p)}(t)$, and first three FPCs, $\phi^{(p)}_k(t)$.}
\label{supp_fig:Multi_Func_Cov}
\end{figure}

From Supplemental Figure~\ref{supp_fig:Multi_Func_Cov}, we find that MSFAST produces near nominal mean coverage for basically all functional components in both data density scenarios. Further, these coverage distributions were rather tight, limiting poor coverage in the worse-case simulations.

We also performed a second set of simulations aimed at assessing computational efficiency, mirroring the timing simulation in Section~\ref{sec:MSFAST_sim} but setting the number of unique observation time points to just $M = 500$. This smaller pool of unique observation times should become saturated well before the number of participants reaches $I = 1000$, demonstrating whether mFPCA and mFACEs have the anticipated sub-linear scaling in $I$. The resulting visualization can be found in Supplemental Figure~\ref{supp_fig:Timing}

\begin{figure}[h]
\centering
\includegraphics[width=12cm]{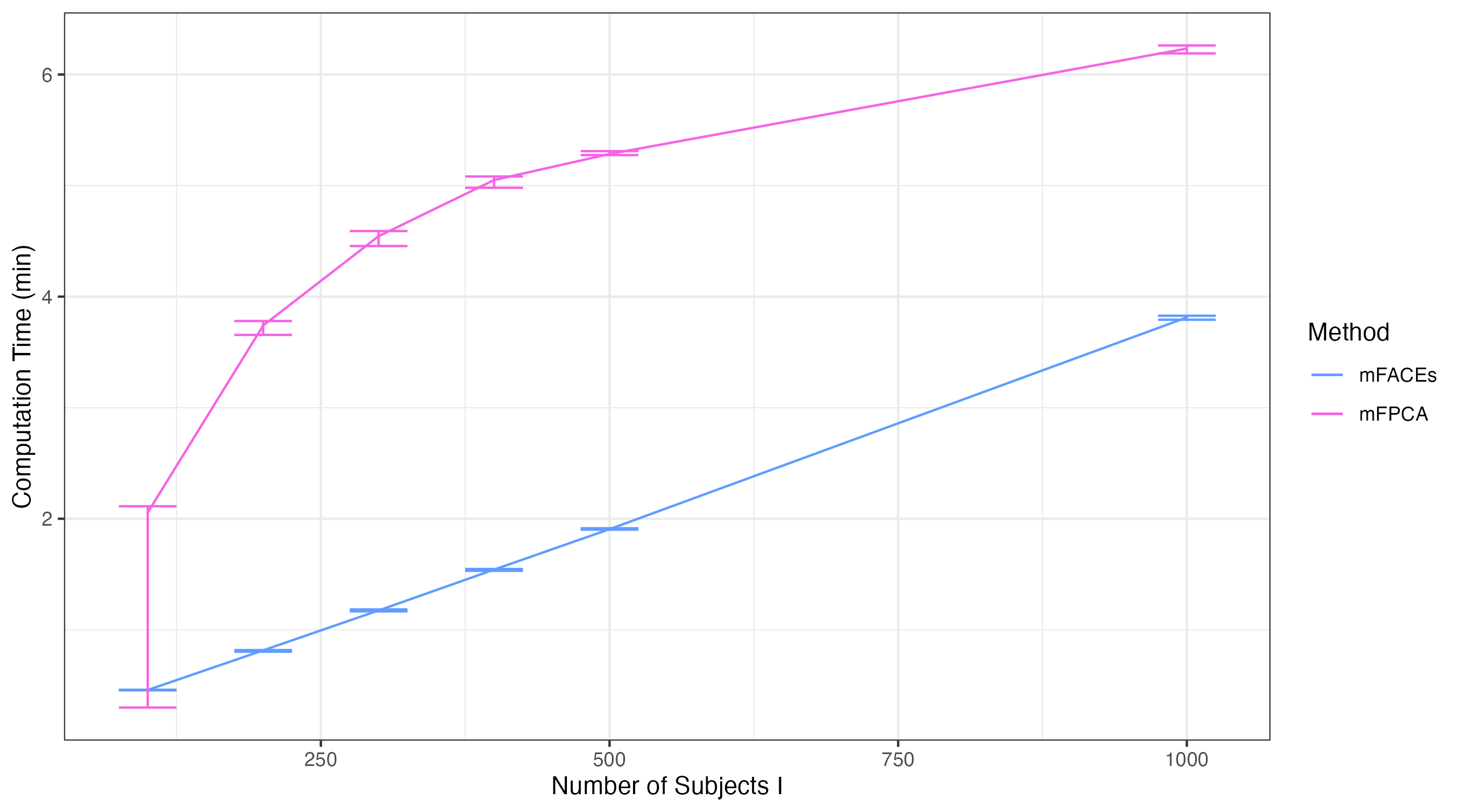}
\caption{Computation time (y-axis in minutes) as a function of number of subjects $I$ (x-axis) for mFACEs, mFPCA, and VMP.
Lines: median time; error bars: min and maximum time.}
\label{supp_fig:Timing}
\end{figure}

From Supplemental Figure~\ref{supp_fig:Timing}, it is clear that mFPCA indeed demonstrates the expected sub-linear scaling. However, mFACEs actually remains linear over this domain. This could be a result of the adjusted prediction procedure (performing predictions one subject at a time) required to ensure that mFACEs does not exhaust available memory. In the absence of performing predictions of the latent trajectories, the expected behavior would likely occur as it did for mFPCA, but more testing is required.

\subsection*{CONTENT Variance Explained}\label{supp:var_exp}

\begin{figure}[h]
\centering
\includegraphics[width=12cm]{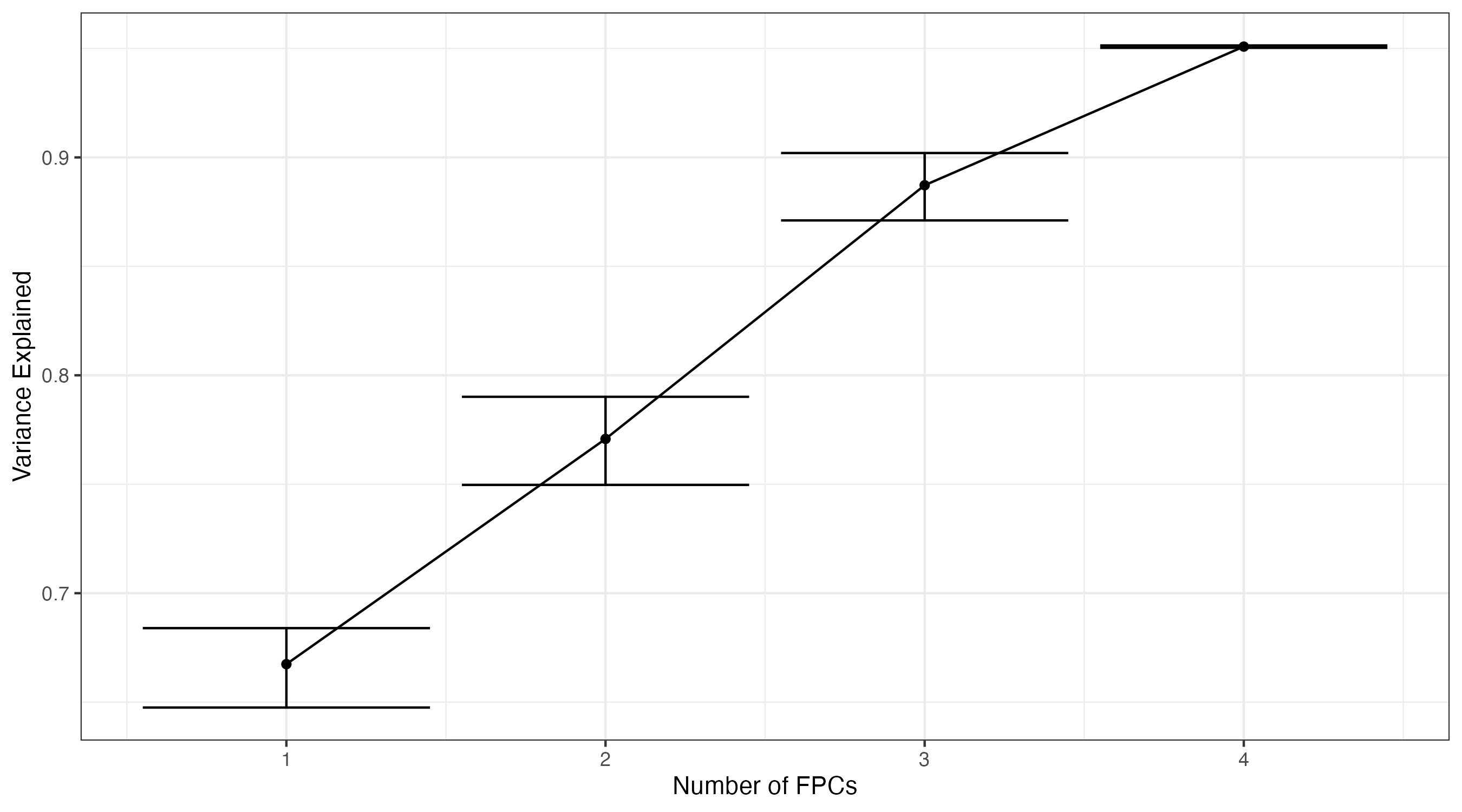}
\caption{Proportion of variance explained in the original data for each truncation number of FPCs $K$. Points represent posterior mean estimates, while error bars correspond to equal-tailed 95\% credible intervals from the posterior samples}
\label{supp_fig:Content_Scree}
\end{figure}

We provide a graph of variance explained at various levels of truncation $K$ in Figure~\ref{supp_fig:Content_Scree}. Of note, we perform the rotational and score alignment procedure discussed in Section~\ref{sec:ACE} prior to the truncation. This ensures that the FPCs are ordered in terms of variance explained, with the first explaining the largest amount. From Figure~\ref{supp_fig:Content_Scree}, we see that using 4 FPCs consistently explains $\approx 95\%$ of the global variability in the multivariate CONTENT data, with truncating further to 3 resulting in only explaining $\approx 88\%$.

\clearpage

\bibliographystyle{abbrvnat}
\bibliography{reference}

\end{document}